\documentclass[a4paper]{article}
\usepackage{caption}
\usepackage{comment}
\usepackage{color}
\usepackage{amsmath}
\usepackage{amssymb}
\allowdisplaybreaks[2]
\newtheorem{theorem}{Theorem}
\newtheorem{prop}{Proposition}

\newtheorem{lem}{Lemma}

\newcommand{\ba}{\begin{eqnarray}}
\newcommand{\ea}{\end{eqnarray}}
\newcommand{\ban}{\begin{eqnarray*}}
\newcommand{\ean}{\end{eqnarray*}}
\newcommand{\no}{\nonumber}

\newcommand{\bc}{\mathbb{C}}
\def\d{{\partial}}
\newcommand{\mapright}[1]{%
\smash{\mathop{%
\hbox to 1.0cm{\rightarrowfill}}\limits^{#1}}}
\newcommand{\mapleft}[1]{%
\smash{\mathop{%
\hbox to 1.3cm{\leftarrowfill}}\limits^{#1}}}

\begin{document}
\title{
\begin{flushright}
  \begin{minipage}[b]{5em}
    \normalsize
    ${}$      \\
  \end{minipage}
\end{flushright}
{\bf Holomorphic Vector Field and Topological Sigma Model on $\mathbb{C}\mathrm{P}^1$ World Sheet}}
\author{Masao Jinzenji${}^{(1)}$, Ken Kuwata${}^{(2)}$\\
\\
(1) \it Department of Mathematics,  \\
\it Okayama University \\
\it  Okayama, 700-8530, Japan\\
\\
(2) \it Division of Mathematics, Graduate School of Science \\
\it Hokkaido University \\
\it  Kita-ku, Sapporo, 060-0810, Japan\\
\\
\it e-mail address: (1) pcj70e4e@okayama-u.ac.jp, \\
\it\hspace{3.5cm}(2) kuwata@math.sci.hokudai.ac.jp }
\maketitle

\begin{abstract}
Witten suggested that fixed-point theorems can be derived by the supersymmetric sigma model on a Riemann manifold $M$ 
with potential terms induced from a Killing vector on $M$ \cite{ref3}.  
One of the well-known fixed-point theorems is the Bott residue formula \cite{ref2} which represents the  intersection number of Chern classes of holomorphic vector bundles on a K$\ddot{\mbox{a}}$hler manifold $M$ as the sum of contributions from fixed point sets of a holomorphic 
vector field $K$ on $M$. 
In this paper, we derive  
the Bott residue formula by using the topological sigma model (A-model) that describes dynamics of maps from $\mathbb{C}P^{1}$ to $M$, with potential terms induced from the vector field $K$. Our strategy is to restrict phase space of path integral to maps homotopic to constant maps.
As an effect of adding a potential term to the topological sigma model, we are forced to modify the BRST symmetry of the original topological sigma model.    
Our potential term and BRST symmetry are closely related to the idea used in the paper by Beasley and Witten \cite{ref12} where potential terms induced from holomorphic section of a holomorphic vector bundle and corresponding supersymmetry are considered. 
\end{abstract}

\section{Introduction}

\subsection{Background}
A supersymmetric transformation is a transformation that exchanges  bosonic fields and fermionic fields in the Lagrangian. If the Lagrangian is invariant under supersymmetric transformation, the model is called  having supersymmetry. Especially,  the sigma model having supersymmetry is called supersymmetric sigma model.
Various topological indices of a Riemann manifold $M$, such as Euler number, Hirzebruch signature, Atiyah-Singer index etc.,  are computed by using a path integral of the supersymmetric sigma model with target space $M$ \cite{ref1, ref14, ref16}. 
In \cite{ref9}, Witten considered the supersymmetric sigma model  with various potential terms. 
Especially, he suggested that the fixed-point formula for signature of even-dimensional $M$ can be obtained by using this model
with a potential term induced from a Killing vector field on $M$.
On the other hand, various fixed-point formulas, such as Duistermaat-Heckman formula etc.,  have been 
derived by using this kind of potential terms \cite{ref4, ref6}.  

Our aim of this paper is to derive the Bott residue formula \cite{ref2}, that describes intersection number of Chern classes of holomorphic vector bundles on a K$\ddot{\mbox{a}}$hler manifold $M$ as the sum of contributions from fixed point sets of a holomorphic 
tangent vector field $K$ on $M$, by using the topological sigma model (A-model) from $\mathbb{C}P^{1}$ to $M$ with  potential terms induced from the vector field $K$. 
It is well-known that the BRST symmetry of the topological sigma model is obtained from the supersymmetric transformation 
of the $N=(2,2)$ supersymmetric sigma model after operating A-twist on the fields of the model.  
In order to extend the BRST-symmetry to the A-model with the potential terms, 
we have to modify the BRST-transformation of the usual topological sigma model. Supersymmetric transformation of  the  $N=(2,2)$ supersymmetric sigma model has four fermionic infinitesimal deformation parameters $\alpha_{+},\alpha_{-},\tilde{\alpha}_{+},\tilde{\alpha}_{-}$.
The BRST-symmetry of the topological 
sigma model (A-model ) is obtained from setting $\alpha_{-}=\tilde{\alpha}_{+}=\alpha,\;\alpha_{+}=\tilde{\alpha}_{-}=0$
after operating A-twist on the fields. 
On the other hand,  our BRST symmetry is obtained from setting $\tilde{\alpha}_{+}=\bar{\alpha},\;\alpha_{-}=\alpha_{+}=\tilde{\alpha}_{-}=0$ after operating A-twist on the fields. 
Hence our BRST symmetry uses half of the supersymmetry used in constructing the usual BRST symmetry of 
the topological sigma model (A-model).    
But our BRST symmetry still has nilpotency, which  will be shown later by explicit computations. The new BRST symmetry is still conserved
in the original Lagrangian of the A-model, and it can be extended to the A-model with the potential term. 
As a result, BRST-closed observables correspond to Dolbeault cohomology of $M$ instead of De Rham cohomology of $M$. But Chern classes of holomorphic 
vector bundles on $M$ are given by $(i,i)$ forms of $M$, and they are automatically  Dolbeault cohomology classes of $M$. 
Therefore, change of BRST-symmetry causes no problem for our purpose in this paper. Potential terms similar to our model have already proposed by Labastida and Llatas in \cite{ref17}. They  construct a supersymmetric sigma model and derived Poincar$\acute{e}$-Hopf theorem. However, we note that their supersymmetry is not BRST-symmetry and their result is different from our aim of this paper. Recently,
Beasley and Witten considered supersymmetry closely related to our new BRST-transformation
for  $(0,2)$ liner sigma models with a potential term induced from holomorphic section of a holomorphic vector bundle \cite{ref12}.
Our new BRST-symmetry seems to be closely related to their idea applied to the case when the holomorphic vector bundle 
is given by  the holomorphic tangent vector bundle $T^{\prime}M$.  Of course, they derive a kind of 
fixed-point formula, but their result is different from our goal in this paper: ``deriving the Bott-residue formula by using the topological sigma model (A-model) with the potential term induced from a holomorphic tangent vector field''.

\subsection{Main result}

 The purpose of this paper is to provide a derivation of the Bott residue formula by using the topological sigma model (A-model)
with potential terms induced from a holomorphic tangent vector field.
First, we introduce assertion of the Bott residue formula \cite{ref2}.\\
\\
{\bf The Bott Residue Formula}
\begin{equation*}
 \int_M \underline{\varphi}(E)= \displaystyle \sum_{\alpha} \int_{N_{\alpha}} \frac{\underline{\varphi}(\Lambda_{\alpha})}
{\det( \theta^\nu_{\alpha} +tR^\nu_{\alpha})} \biggl|_{t=\frac{ i }{2 \pi }}.
\end{equation*}
Here, $M$ is a compact K$\ddot{\mbox{a}}$hler manifold with $\mathrm{dim}_{\mathbb{C}}(M)=m$. Let $E$ be a holomorphic vector bundle over $M$ with $\mathrm{rank}_{\mathbb{C} } E=q$ and 
$\underline{\varphi}(E)$ is a wedge product of  Chern classes of $E$ represented by symmetric homogeneous polynomial 
$\varphi(x_{1},\cdots,x_{q})$ of degree $m$ (explicit definition will be introduced later). 
Let $K$ be a holomorphic tangent vector field on $M$ and $\{N_{\alpha}\}$ be set of connected components of the zero set of $K$. 
$\underline{\varphi}(\Lambda_{\alpha})$ is a cohomology class of $N_{\alpha}$ which will be defined  later.
$\theta^\nu_{\alpha}$ is the automorphism induced by action of $K$ on the normal bundle of $N_{\alpha}$. $R^\nu_{\alpha}$ is the curvature $(1,1)$-form of the normal bundle of $N_{\alpha}$. 

Next, we  introduce the topological sigma model (A-model). We use the 
topological sigma model (A-model) from $\mathbb{C}P^1$ to the K$\ddot{\mbox{a}}$hler manifold $M$ with potential terms induced from the holomorphic 
tangent vector field $K=K^{i}\frac{\d}{\d z^{i}}$. 
Fields that appear in the model is given as follows.
\begin{itemize}
\item{$\phi^{i}$ $\phi^{\bar{i}}$ : bosonic fields given as $C^\infty$-map from $\mathbb{C}P^{1}$ to $M$ }
\item{$\chi^{i}$:  fermionic fields that takes values in $C^\infty$ section of $\phi^{-1} T^{\prime} M$}
\item{$\chi^{\bar{i}}$:  fermionic fields that takes values in $C^\infty$ section of $\phi^{-1} \overline{T^{\prime}} M$ }
\item{$\psi^{\bar{i}}_z$: fermionic fields that takes values in $C^\infty$ section of $T^{\prime *}\mathbb{C}P^{1} \otimes \phi^{-1} \overline{T^{\prime}}M$ }
\item{$\psi^{i}_{\bar z}$: fermionic fields that takes values in $C^\infty$ section of
$\overline{T^{\prime *}}\mathbb{C}P^{1}\otimes \phi^{-1} T^{\prime}M $}
\end{itemize}
Let $g_{i \bar j}$ be  K$\ddot{\mbox{a}}$hler metric of $M$ and $R_{i \bar j k \bar l}$ be its curvature tensor. 
Then the Lagrangian of our model is given as follows \cite{ref5}.
\begin{align}
L+V= \int_{\mathbb{C} \mathrm{P^1}} dz d\bar{z} \bigl[\frac{t}{2} g_{i \bar{j} } ( \partial_z \phi^i \partial_{\bar z}\phi^{\bar j} +  \partial_{\bar z} \phi^i \partial_z \phi^{\bar j}) +  \sqrt{t}i g_{i \bar{j}} \psi^{\bar{j}}_z  D_{\bar z}  \chi^i +\sqrt{t}ig_{i \bar{j}} {\psi}^i_{\bar z} D_z  \chi^{\bar{j}} \notag \\- R_{i \bar{j} k \bar{l}} {\psi}^i_{\bar z} \psi^{\bar{j}}_z \chi^{k} \chi^{\bar l} +ts^2\beta g_{i \bar{j}} K^i \bar{K}^{\bar{j}} + t sg_{i \bar{j}} \triangledown_{\bar{\mu}} \bar{K}^{\bar{j}} \chi^{\bar {\mu}} \chi^i + s\beta g_{i \bar{j}} \triangledown_{\mu} K^{i}  \psi^{\mu}_{\bar{z}}  \psi^{\bar{j}}_z \bigr]
\end{align}
From now on, we set $\beta:=2\pi i$.  Covariant derivatives are given by
\begin{equation}
D_{\bar z}  \chi^i =\partial_{\bar z} \chi^i +\Gamma^i_{\mu \nu} \partial_{\bar z} \phi^\mu \chi^\nu,
\end{equation}
\begin{equation}
D_{ z}  \chi^{\bar i} =\partial_z \chi^{\bar i} +\Gamma^{\bar i}_{\bar \mu \bar \nu} \partial_z \phi^{\bar \mu} \chi^{\bar \nu}.
\end{equation}
Our BRST-transformation for this model is given as follows. ($\bar \alpha$ is a fermionic variable.)
\begin{align}
\delta \phi^{\bar i} &=i \bar{\alpha} \chi^{\bar i}, &
\delta \psi^{\bar{i}}_z &=-i \bar{\alpha} \Gamma^{\bar i}_{\bar {\mu} \bar{\nu}}\chi^{\bar{\mu}} \psi^{\bar{\nu}}_z, & \notag \\
\delta{\psi}^i_{\bar z} &=- \sqrt{ t} \bar{\alpha} \partial_{\bar z} \phi^i , &  \delta \chi^i &= i s\bar \alpha \beta K^i,
&\delta{\phi}^i   =\delta \chi^{\bar i} =0.  \label{a1}
\end{align}
The above potential terms and BRST-symmetry are closely related to the supersymmetry used in \cite{ref12}. 
In this paper, we prove the following proposition that play important roles in our derivation.
\begin{prop} 
Correlation functions of BRST-closed observables are  invariant under variation of $s$.
\label{p1}
\end{prop}
Then we can derive the Bott residue formula by evaluating correlation function of degree zero map
both in the limit $s\rightarrow 0$ and $s\rightarrow \infty$.

\subsection{Organization of the Paper}
This paper is organized as follows.

 In section 2, we introduce the Bott residue formula and notations used in this paper.  And to demonstrate usage of the Bott resideuformula, we compute the integral of Chern classes of a hyperplane bundle on $\bc P^n$.

In section 3, we introduce our topological sigma model. First, we review outline of the topological sigma model (A-model) without potential terms and introduce our BRST-symmetry that uses half of the supersymmetry used in the usual BRST-symmetry of the 
original topological sigma model. 
We show that our new BRST-symmetry conserves the Lagrangian of the topological sigma model without potential terms.
Under the new BRST-symmetry, BRST-closed observables become elements of Dolbeault cohomology of the target K$\ddot{\mbox{a}}$hler manifold. 
Next, we include potential terms induced from a holomorphic tangent vector field and extend the new BRST-symmetry to this case. 
Then BRST-closed observables become elements of equivariant Dolbeault cohomology under the action of the holomorphic vector field. Mathematical relationship between the Bott residue formula \cite{ref9} and this cohomology is discussed in \cite{ref15, ref13}.

Section 4 is the main section of this paper. We evaluate the degree $0$ (i.e., homotopic to constant maps) correlation function of our model. It corresponds to the correlation function represented by the Bott residue formula. 
Proposition \ref{p1} claims that the correlation function is invariant under change of the parameter $s$. Hence we can compare  the 
results evaluated under the $s \to 0$ limit and the $s \to \infty$ limit. 

In the $s \to 0$ limit, the degree $0$ correlation function turns into the classical integration on $M$ of differential forms that represent Chen classes by the standard argument of weak coupling limit of the topological sigma model. 
In the $s \to \infty$ limit, evaluation of the degree $0$ correlation function reduces to the evaluation of  contributions of from connected components of the zero set of the holomorphic tangent vector field $K$.  This follows from the localization principle. The localization principle insists that path integrals of a model having supersymmetry can be evaluated by the sum of  contributions from field configurations satisfying the conditions that supersymmetric transformations of fermionic fields are zero.
We perform standard localization computation. The result of evaluation from each connected component turns out to be the contribution in the Bott residue formula from the same connected component.
By equating these two results, we obtain the desired Bott residue formula.

In Appendix, we prove Proposition $1$ used in our derivation of the Bott Residue Formula.

\section{Introducton to the Bott residue formula}
\subsection{Notation and The Bott Residue Formula}\label{kigou}

 We introduce here our basic notations. 
 Let us denote  a compact K$\ddot{\mbox{a}}$hler manifold  by $M$ ($\mathrm{dim}_{\mathbb{C}}(M)=m$) and  a holomorphic tangent vector field on $M$ by $K$. Let $E$ be a holomorphic vector bundle on $M$ with $\mathrm{rank}_{\mathbb{C} } (E)=q$ and $\Gamma (E) $ be set of $C^\infty $ sections of $E$.

 First, we note the basic facts of $E$ and introduce an action of $K$ on $E$ used in \cite{ref2}.
 The holomorphic vector bundle $E$ has a canonical connection compatible with Hermitian metric. 
Let $\tilde{\nabla}$ be the canonical connection on $E$ and $\Omega^{p,r}(E)$ be complex vector space of 
$E$-valued $(p,r)$-forms.
We also introduce the exterior holomorphic covariant derivative $D': \Omega^{p,r}(E) \to \Omega^{p+1,r}(E)$.
Then the canonical connection is decomposed into  $\tilde{\nabla}=D'+\bar \partial$ where $\bar \partial$ is the anti-holomorphic part of the exterior derivative operator $d$ of $M$.
Let $\{e_a\;|\; a=1,\cdots, q\}$ be local holomorphic frame of $E$. Then the following relation holds.
\begin{align}
&\bar \partial e_a=0,\no\\
&\tilde{\nabla}e_a =D'e_a=\Theta^b_{ a k }dz^{k} e_b ,
\end{align}
where $(z^1,\cdots,z^m)$ is a local coordinate of $M$ and $\Theta^b_{ a k }dz^{k}$ is the connection $(1,0)$-form of $E$.
Curvature $(1,1)$-form $F=(F^{b}_{a}),F^{b}_{a}=F^{b}_{ak\bar{l}}dz^{k}\wedge d\overline{z^{l}},$ is given by $F^{b}_{a}=\bar{\d}(\Theta^b_{ a k }dz^{k})
=-\bar{\d}_{l}(\Theta^b_{ a k })dz^{k}\wedge d\overline{z^{l}}$. 
Here, we define $\Lambda :\Gamma (E) \to \Gamma (E) $ as a  differential operator which acts on $fs$ ($f:\; C^{\infty}$ a function on $M$, 
$s\in \Gamma(E)$) in the following way,
\begin{align}
\Lambda (fs) &= (Kf) \cdot s + f \Lambda (s), & \bar \partial \Lambda & = \Lambda \bar \partial. &
\label{lcond}
\end{align}
This $\Lambda$ defines the action of $K$ on $E$.
In the case when $E$ is the holomorphic tangent bundle $T^{\prime}M$,  $\Lambda$ is explicitly given by holomorphic Lie derivative of the holomorphic tangent vector field $Y$ by $K$:
\ba
\theta(K) : Y \to [K,Y].
\ea 
Then, we can check that $\theta(K)$ satisfies the condition (\ref{lcond}) by using local coordinate on $M$. 
\begin{align}
& \theta(K)(fY)=[K,fY]=K^i\frac{\partial}{\partial z^i}(fY^j \frac{\partial}{\partial z^j})-fY^j \frac{\partial}{\partial z^j}(K^i\frac{\partial}{\partial z^i}) \nonumber\\
&=K^i\frac{\partial f}{\partial z^i}(Y^j \frac{\partial}{\partial z^j})+fK^i\frac{\partial}{\partial z^i}(Y^j \frac{\partial}{\partial z^j})-fY^j \frac{\partial}{\partial z^j}(K^i\frac{\partial}{\partial z^i}) \nonumber\\
&=K^i\frac{\partial f}{\partial z^i}(Y^j \frac{\partial}{\partial z^j})+f[K,Y] \nonumber\\
&=(Kf)Y+f[K,Y]. \label{thetacondition}
\end{align}

Second, we introduce notations for characteristic classes of $E$.
Let $\varphi(x_1 , \cdots ,x_q)$ be a symmetric homogeneous polynomial in $x_1,\cdots,x_{q}$ with complex coefficients of homogeneous 
degree $m=\mathrm{dim}_{\mathbb{C}}(M)$.
We define $\varphi(A)$ where $A$ is an endomorphism $A:V \to V$. ($V$: a complex $q$-dimensional vector space ).
Let $\lambda_i$ $(I=1,\cdots q)$ be eigenvalues of $A$. Then it is defined by
 \begin{equation}
\varphi(A) := \varphi(\lambda_1 , \cdots , \lambda_q).
\label{A}
\end{equation}
We then regard $x_1,\cdots,x_{q}$ as Chern roots of $E$ defined by
\begin{equation}
\prod _{i=1}^q(1+tx_i):=1+tc_1(E)+t^2c_2(E)+\cdots +t^qc_q(E).
\end{equation}
With this set-up, a characteristic class $\underline{\varphi}(E)$ is defined as follows.
\ba
\underline{\varphi}(E):=\varphi(x_1 , \cdots ,x_q).
\ea

Let $\{N_{\alpha}\}$ be the set of connected components of the zero set of $K$.
We assume that each $N_{\alpha}$ is a compact K$\ddot{\mbox{a}}$hler submanifold of $M$.
In the following, we define $\underline{\varphi}(\Lambda_{\alpha})$, which is given as a cohomology class of $N_{\alpha}$.
Let $\Lambda_{\alpha}$ be $\Lambda|_{N_{\alpha}}$, i.e., restriction of $\Lambda$ to $N_{\alpha}$.
By the first condition in (\ref{lcond}), $\Lambda_{\alpha}$ becomes an endmorphism of $E_{\alpha}:=E|_{N_{\alpha}}$. 
We say that $\Lambda$ is constant type if eigenvalues of $\Lambda_{\alpha}$ are constant on each  connected component $N_{\alpha}$. 
In this paper, we assume  that $\Lambda$ is constant type.
Then we introduce the following notations.
 Let $\{ \lambda_i^{\alpha}\;|\;i=1,\cdots, r \}$ be distinct eigenvalues of $\Lambda_{\alpha}$ $(r\leq q)$ and
$m_{i}^{\alpha}$ be multiplicity of  $\lambda_{i}^{\alpha}$ ($\sum_{i=1}^{r}m_{i}^{\alpha}=q$).
We denote the largest sub-bundle of $E_{\alpha}$ on which $(\Lambda_{\alpha} -\lambda_i^{\alpha})$ is nilpotent by $E_{\alpha}(\lambda_i^{\alpha})$ 
($\mathrm{rank}_{\mathbb{C}}(E_{\alpha}(\lambda_i^{\alpha}))=m_{i}^{\alpha}$).
Then, $E_{\alpha}$ canonically decomposes into a direct sum, 
$\displaystyle{E_{\alpha}=\mathop{\oplus}_{i=1}^{r}E_{\alpha}(\lambda_i^{\alpha})}$.
Let $c_i(E_{\alpha}(\lambda_{i}^{\alpha}))$ be the $i$-th Chern class of $E_{\alpha}(\lambda_{i}^{\alpha})$
and $x_j(\lambda_{i}^{\alpha})\;\;(j=1,\cdots,m_{i}^{\alpha})$ be Chern roots of $E_{\alpha}(\lambda_{i}^{\alpha})$ defined by
\begin{equation}
\prod _{j=1}^{m_{i}^{\alpha}}(1+tx_j(\lambda_{i}^{\alpha})):=1+tc_1(E_{\alpha}(\lambda_{i}^{\alpha}))+t^2c_2(E_{\alpha}(\lambda_{i}^{\alpha}))+\cdots +t^qc_q(E_{\alpha}(\lambda_{i}^{\alpha})).
\end{equation}
With these set-up's,  the cohomology class $\underline{\varphi}(\Lambda_{\alpha})$ is defined by
\ba
\underline{\varphi}(\Lambda_{\alpha})&:=&\varphi(\lambda_1^{\alpha} +x_1(\lambda_{1}^{\alpha}) ,\cdots,
\lambda_1^{\alpha} +x_{m_{1}^{\alpha}}(\lambda_{1}^{\alpha}),\lambda_2^{\alpha} +x_1(\lambda_{2}^{\alpha}) ,\cdots,
\lambda_2^{\alpha} +x_{m_{2}^{\alpha}}(\lambda_{2}^{\alpha}),\no\\
&&\cdots,\lambda_r^{\alpha} +x_1(\lambda_{r}^{\alpha}) ,\cdots,\lambda_r^{\alpha} +x_{m_{r}^{\alpha}}(\lambda_{r}^{\alpha})).
\ea
This is the original definition of $\underline{\varphi}(\Lambda_{\alpha})$ used in \cite{ref9}. Let $F_{\alpha}$ be curvature
$(1,1)$-form (valued in $\mathrm{End}(E|_{N_{\alpha}})$) of $E|_{N_{\alpha}}$ If we regard $\Lambda_{\alpha}+\frac{i}{2\pi}F_{\alpha}$
as $\mathrm{End}(E|_{N_{\alpha}})$ valued form on $N_{\alpha}$, $\underline{\varphi}(\Lambda_{\alpha})$ is rewritten by using 
(\ref{A}) as follows.
\ba
\underline{\varphi}(\Lambda_{\alpha})=\varphi(\Lambda_{\alpha}+\frac{i}{2\pi}F_{\alpha}).
\ea   
With these set-up's, we introduce the Bott residue formula. 
\noindent
\noindent
We assume that the endomorphism $\theta|_{N_{\alpha}}$, induced by the action of $K$ on the holomorphic tangent bundle $T^{\prime}M|_{N_{\alpha}}$ has precisely $T^{\prime}N_{\alpha}$ for its kernel; i.e., the sequence
\begin{align}
0 \rightarrow T^{\prime}N_{\alpha} \rightarrow T^{\prime}M|_{N_{\alpha}} \xrightarrow{\theta |_{N_{\alpha}}} T^{\prime}M|_{N_{\alpha}}
\end{align}
is exact.
From the above exact sequence, $\mathrm{Im}(\theta |_{N_{\alpha}}) \cong T^{\prime}M|_{N_{\alpha}} /T^{\prime}N_{\alpha}$ follows. 
 Hence we obtain an automorphism $\theta^\nu_{\alpha}:=\theta^{\nu}|_{N_{\alpha}} : T^{\prime}M|_{N_{\alpha}} /T^{\prime}N_{\alpha} \to T^{\prime}M|_{N_{\alpha}} /T^{\prime}N_{\alpha} $ (where the subscript ``$\nu$'' means that we consider holomorphic normal bundle 
$T^{\prime}M|_{N_{\alpha}} /T^{\prime}N_{\alpha}$ instead of $T^{\prime}M|_{N_{\alpha}}$ ). 
Let $R^\nu_{\alpha}$ be the curvature (1,1)-form of the holomorphic normal bundle $T^{\prime}M|_{N_{\alpha}} /T^{\prime}N_{\alpha}$
on $N_{\alpha}$.
Then the Bott residue formula is given as follows.
\ba
 \int_M \underline{\varphi}(E)&=& \sum_{\alpha} \int_{N_{\alpha}} \frac{\underline{\varphi}(\Lambda_{\alpha})}
{\det( \theta^\nu_{\alpha} +\frac{i}{2\pi}R^\nu_{\alpha})}\no\\ 
&=&\sum_{\alpha} \int_{N_{\alpha}} \frac{\varphi(\Lambda_{\alpha}+tF_{\alpha})}
{\det( \theta^\nu_{\alpha} +tR^\nu_{\alpha})}\biggl|_{t=\frac{ i }{2 \pi }}.
\label{goal}
\ea
In Section 3 and Section 4, we will derive the Bott Residue formula in the form of the second line of the above equality.

\subsection{Usage of the Bott residue formula: An Example}
\textcolor{red}{}In order to demonstrate usage of the Bott residue formula, we compute the following integral of Chern 
classes by using the formula (\ref{goal}):
\begin{align}
&\int_{\mathbb{C}P^n}c_1(H)^n.
\end{align}
Here, $H$ is a hyperplane bundle on $\mathbb{C}P^n$.

\textcolor{red}{}First, we explain the case of  $\mathbb{C}P^1$. Let $(X_1:X_2)$ be homogeneous coordinates of  $\bc P^1$. 
Then, $\bc P^1$ is covered by two open sets $U_1:=\{ (X_1:X_2) \in \bc P^1|X_1\neq 0\}$ and 
$U_2:=\{(X_1:X_2) \in \bc P^1|X_2 \neq 0 \}$. 
We use $z:=\frac{X_2}{X_1}$ and $w:=\frac{X_1}{X_2}$ as local holomorphic coordinates on $U_1$ and $U_2$ respectively. 
Then we introduce the following holomorphic vector field $K$ on $\bc P^1$.
\begin{align}
K&=\alpha_1 z \frac{d}{d z}& (\text{on } U_1), \\
&=-\alpha_1 w \frac{d}{d w }& (\text{on } U_2),
\end{align}
where we assume $\alpha_{1}\neq 0$.
Then actions of $\theta(K)$ on $\displaystyle{\frac{d}{d z}}$ and $\displaystyle{\frac{d}{d w}}$ are given by
\begin{align}
\theta(K)(\frac{d}{d z})&:=[K,\frac{d}{d z}]=[\alpha_1 z \frac{d}{d z},\frac{d}{d z}]=-\alpha_1\frac{d}{d z},&
\theta(K)(\frac{d}{d w})&=\alpha_1\frac{d}{d w}.&
\label{Ka1}
\end{align}
Zero set of $K$ is given by $\{(1:0),(0:1)\}$ and we denote $(1:0)\;(\leftrightarrow\;z=0)$ and $(0:1)\;(\leftrightarrow\;w=0)$ by $p_{1}$ and $p_{2}$
respectively. Then (\ref{Ka1}) tells us that $\theta^{\nu}|_{p_{1}}=-\alpha_{1}$ and $\theta^{\nu}|_{p_{2}}=\alpha_{1}$
because $T^{\prime}\bc P^{1}|_{p_{i}}/T^{\prime}p_{i}=T^{\prime}\bc P^{1}|_{p_{i}}$. 

Next, we construct local frame of $H=S^*$ in order to determine $\Lambda_{1}=\Lambda|_{p_{1}}$ and $\Lambda_{2}=\Lambda|_{p_{2}}$ ($S$ is the tautological line bundle on $\bc P^1$). 
The fiber of $S$ on $(X_1:X_2) \in \bc P^1$ is given by a complex $1$-dimensional linear subspace of $\bc^2$ spanned by $(X_1,X_2)$.  Let $e_1$ and $e_2$ be local frames on $U_1$ and $U_2$ given by
\begin{align}
&e_1=(1,z),&e_2=(w,1).
\end{align}
Hence transition functions of $S$ that satisfy $e_{\alpha}=g^{S}_{\alpha\beta}e_{\beta}$ on $U_{1}\cap U_{2}$ 
are obtained as $\displaystyle{g^S_{12}=\frac{1}{w}}$ and $\displaystyle{g^S_{21}=\frac{1}{z}}$. Let $f_1$ and $f_2$ be local frames of $H$ on $U_1$ and $U_2$ respectively. Since hyperplane bundle $H$ is dual line bundle of $S$, transition functions of $H$ are given as  $g^H_{12}=(g^S_{12})^{-1}=w$ and $g^H_{21}=z$. Then a global holomorphic section $s$ is given by $f_1$ on $U_{1}$ and 
$wf_2$ on $U_{2}$ because $f_1=wf_{2}$ holds on $U_{1}\cap U_{2}$.
With this set-up, note that $\Lambda$ satisfies (\ref{lcond}) and that $\Lambda(s^1f_1)=\Lambda(s^2f_2)$ on $U_1 \cap U_2$. 
Since $p_{1}=(1:0)\in U_1\setminus U_{2}$ and $p_{2}=(0:1)\in U_2\setminus U_{1}$, 
we set $\Lambda(f_1)=:\Lambda_{1}f_1$ and $\Lambda(f_2)=:\Lambda_{2}f_2$.
Then we apply (\ref{lcond}) to the above global holomorphic section $s$. 
\begin{align}
\Lambda(f_1)&=\Lambda(f_1)=\Lambda_1 f_1, \\
\Lambda(wf_2)&=K(w)f_2+w\Lambda(f_2)
=-\alpha_1w \frac{dw}{dw}f_2+w \Lambda_2 f_2
=(-\alpha_1+\Lambda_2)wf_2.
\end{align}
Since $\Lambda(f_1)=\Lambda(wf_2)$ holds on $U_{1}\cap U_{2}$,
we obtain
\begin{align}
\Lambda_1=C,\;\; \Lambda_2=C+\alpha_1, (\mbox{$C$ is an arbitrary constant}).
\end{align}
With these set-up's, we compute $\int_{\mathbb{C}P^1} c_1(H)$. Since zero set of $K$ is given by $\{p_1,\; p_2\}$ and $c_1(H)$ is given by trace of curvature form of $H$, the assertion of Bott Residue Formula in this case becomes
\begin{align}
\int_{\mathbb{C}P^1} c_1(H)=\sum_{i=1}^2 \frac{\mathrm{tr}(\Lambda|_{p_i})}{\mathrm{det}(\theta^{\nu}|_{p_i})}.
\end{align} 
On the other hand, we have the following table from the discussions so far:
\begin{table}[h]
\caption{\bf Summary of the case of $\bc P^{1}$}
\label{T1}
\centering
\begin{tabular}{|c|c|c|} \hline
&$p_1$&$p_2$\\ \hline
$\mathrm{tr}(\Lambda|_{p_i})$&$C$&$C+\alpha_1$\\ \hline
$\mathrm{det}(\theta^{\nu}|_{p_i})$&$-\alpha_1$&$\alpha_1$ \\ \hline
\end{tabular}
\end{table}

\textcolor{red}{}Hence we obtain
\begin{align}
\int_{\mathbb{C}P^1} c_1(H)=-\frac{C}{\alpha_1}+\frac{C+\alpha_1}{\alpha_1}=1.
\end{align} 

Then we turn into the $\bc P^{n}$ case. Let $(X_1:\cdots :X_{n+1})$ be  homogeneous coordinates of $\bc P^{n}$. Then $\bc P^n$ is covered by $(n+1)$ open sets $U_i:=\{(X_1:\cdots:X_{n+1}) \in \bc P^n\;|\; X_i \neq 0\}\;\;(i=1,2,\cdots,n+1)$. Local coordinate
systems on $U_{i}$: $\phi_i:U_i\to \bc^n$ are defined in the following way.
\begin{align}
\phi_i(X_1:\cdots:X_{n+1})=\Bigl(\frac{X_1}{X_i},\cdots,\frac{X_{i-1}}{X_i},\frac{X_{i+1}}{X_i},\cdots,\frac{X_{n+1}}{X_i}\Bigr)=(z_{(i)}^1,\cdots,z_{(i)}^n).
\end{align}
Coordinate transformations between $U_1$ and $U_j (j=2,3,\cdots,n+1)$ are explicitly given by
\begin{align}
&z^1_{(j)}=\frac{1}{z_{(1)}^{j-1}},& &z^i_{(j)}=\frac{z_{(1)}^{i-1}}{z_{(1)}^{j-1}} ~~~~(1<i\leq j-1),&
&z^i_{(j)}=\frac{z_{(1)}^{i}}{z_{(1)}^{j-1}} ~~~~(j \leq i).&
\end{align}
From these results, we obtain the following transformation rules on $U_{1}\cap U_{i}$:
\begin{align}
\frac{\d }{\d z^{j-1}_{(1)}}&=\frac{-1}{\d z^{j-1}_{(1)}} \sum_{l=1}^n z_{(j)}^l \frac{\d }{\d z^{l}_{(j)}}, \\
\frac{\d }{\d z^{i}_{(1)}}&=\frac{z_{(j)}^{i+1}}{\d z^{i}_{(1)}} \frac{\d }{\d z^{i+1}_{(j)}} ~~~~(1\leq i\leq j-2), \\
\frac{\d }{\d z^{i}_{(1)}}&=\frac{z_{(j)}^{i}}{\d z^{i}_{(1)}} \frac{\d }{\d z^{i}_{(j)}} ~~~~(j\leq i).
\end{align}
Then, we introduce the following holomorphic tangent vector field $K$ on $\bc P^{n}$:
\begin{align}
K&=\sum_{i=1}^n \alpha_i z_{(1)}^i \frac{\d}{\d z_{(1)}^i}\quad\bigl(\text{on}\;\; U_1\bigr) \\
&=-\alpha_{j-1} z^1_{(j)}\frac{\d}{\d z^1_{(j)}}
+\sum_{i=2}^{j-1} (\alpha_{i-1}-\alpha^{j-1})z^i_{(j)}\frac{\d}{\d z^i_{(j)}} \notag \\
&+\sum_{i=j}^{n} (\alpha_{i}-\alpha_{j-1})z^i_{(j)}\frac{\d}{\d z^i_{(j)}} \quad  \bigl(\text{on}\;\; U_i\;\;(i=2,3,\cdots,n+1)\bigr),
\end{align}
where we assume $\alpha_i\neq 0\;(i=1,2,\cdots,n)$ and $\alpha^i\neq \alpha^j\;\;(i\neq j)$. Zero set of $K$ is given by
\begin{eqnarray}
&&\{p_{1},p_{2},\cdots, p_{n+1}\},\no\\
&&\bigl(p_{i}:=\overbrace{0:\cdots:0}^{i-1}:1:\overbrace{0:\cdots:0}^{n-i+1})\;\; (i=1,2,\cdots ,n+1)\bigr).
\end{eqnarray}
Then we determine action of $\theta(K)$ on each $U_{i}$. On $U_1$, it is given by 
\begin{align}
\theta(K)(\frac{\d}{\d z^i_{(1)}})&=-\alpha^i \frac{\d}{\d z^i_{(1)}}.
\end{align}
On $U_j\;\; (j=2,3,\cdots,n+1)$, it is given as follows. 
\begin{align}
\theta(K)(\frac{\d}{\d z^1_{(j)}})&=\alpha^{j-1} \frac{\d}{\d z^1_{(j)}},\notag\\
 \theta(K)(\frac{\d}{\d z^i_{(j)}})&=(\alpha^{j-1}-\alpha^{i-1})\frac{\d}{\d z^i_{(j)}} \;\;(i=2,3,\cdots,j-1),\notag \\
\theta(K)(\frac{\d}{\d z^i_{(j)}})&=(\alpha^{j-1}-\alpha^i)\frac{\d}{\d z^i_{(j)}}\;\;(i=j,j+1,\cdots,n+1)  .
\end{align}
Since $p_{i}\;(i=1,2,\cdots,n+1)$ is a point in $\bc P^{n}$ and $p_{i}\in U_{i}\setminus (\mathop{\cup}_{j\neq i}U_{j})$, $\theta_{i}^{\nu}
:=\theta^{\nu}|_{p_{i}}$ is given by the above representations of $\theta(K)$ on $U_{i}$.
  
Next, we construct local frames of hyperplane bundle $H=S^*$ on $U_{i}$ in order to determine $\Lambda_{i}:=\Lambda|_{p_{i}}$. 
In the same way as the $\bc P^{1}$ case, the local frame of $S$ on $U_i$ is given by
\begin{align}
e_i=(z_{(i)}^1,\cdots,z_{(i)}^{i-1},1,z_{(i)}^i,\cdots,z_{(i)}^n).
\end{align} 
Then we can determine transition function $g_{1j}^{S}$ that satisfies $e_{1}=g_{1j}^{S}e_{j}$ on $U_{1}\cap U_{j}$.
\begin{align}
e_1&=(1,z_{(1)}^1,\cdots,z_{(1)}^n)=z_{(1)}^{j-1}\Bigl(\frac{1}{z_{(1)}^{j-1}},\frac{z_{(1)}^{1}}{z_{(1)}^{j-1}},\cdots,\frac{z_{(1)}^{j-2}}{z_{(1)}^{j-1}},1,\frac{z_{(1)}^{j}}{z_{(1)}^{j-1}},\cdots,\frac{z_{(1)}^{n}}{z_{(1)}^{j-1}} \Bigr) \notag \\
&=z_{(1)}^{j-1} e_j :=g_{1 j}^S e_j~~~(j=2,3,\cdots,n+1).
\end{align} 
Let $f_i (i=1,\cdots,n+1)$ be the local frame of $H=S^{*}$ on $U_i$. Then transition functions $g_{1j}^H$ is given by $(g_{1 j}^S)^{-1}=\frac{1}{z_{(1)}^{j-1}}=z_{(j)}^1$ and we have $ f_1=g_{1 j} f_j =z_{(j)}^1f_j $.
Then we can construct a global holomorphic section $s$ of $H$ which is reperesented as $s^i f_i$ on $U_{i}$, 
by setting $s^1=1,\;\; s^j=z_{(j)}^1\;\; (j=2,3,\cdots,n+1)$
Since $p_{i}\in U_{i}\setminus(\mathop{\cup}_{j\neq i}U_{j})$. we can set $\Lambda(f_i)=\Lambda_i f_i$. 
Then by applying (\ref{lcond}) to $s$, the following equality:  
\begin{align}
&\Lambda(s^1f_1)=\Lambda_1f_1\notag\\ 
&=\Lambda(s^if_i)=K(z_{(i)}^1 )f_i+z_{(i)}^1 \Lambda(f_i)
=-\alpha^{i-1}z_{(i)}^1 f_i+z_{(i)}^1 \Lambda_{i} f_i
=(-\alpha^{i-1}+ \Lambda_{i})f_1,\notag
\end{align}
holds on $U_{1}\cap U_{i}$.
Hence $\Lambda_{i}$'s are given as $\Lambda^1=C,\:\;\Lambda^i=C+\alpha^{i-1} \;\;(i=2,3,\cdots,n+1)$
where $C$ is an arbitrary constant.
We summarize the above results in the following table.
\begin{table}[hb]
\caption{\bf Summary of the case of $\bc P^{n}$}
\label{T2}
\centering
\begin{tabular}{|c|c|c|} \hline
&$p_1$&$p_i\;\;(i=2,\cdots,n+1)$\\ \hline
$\mathrm{tr}(\Lambda|_{p_i})$&$C$&$C+\alpha^{i-1}$\\ \hline
$\mathrm{det}(\theta^{\nu}|_{p_i})$&$\prod_{i=1}^n(-\alpha^i)$&$\alpha^{i-1}\prod_{j=2}^{i-1}(\alpha^{i-1}-\alpha^{j-1})\prod_{j=i}^n(\alpha^{i-1}-\alpha^j)$ \\ \hline
\end{tabular}
\end{table}

Then the assertion of Bott residue formula is given by
\begin{align}
\int_{\mathbb{C}P^n}(c_1(H))^n&=\sum_{i=1}^{n+1} \frac{(\mathrm{tr}(\Lambda|_{p_i})^n}{\mathrm{det}(\theta^{\nu}|_{p_i})} \notag \\
&=\frac{C^n}{\prod_{i=1}^n(-\alpha^i)}
+\sum_{i=2}^{n+1} \frac{(C+\alpha^{i-1})^n}{\alpha^{i-1}\prod_{j=1}^{i-1}(\alpha^{i-1}-\alpha^{j-1})\prod_{j=i}^n(\alpha^{i-1}-\alpha^j)} \notag \\
&=\frac{C^n}{\prod_{i=1}^n(-\alpha^i)}
+\sum_{i=1}^{n} \frac{(C+\alpha^{i})^n}{\alpha^{i}\prod_{j=1}^{i}(\alpha^{i}-\alpha^{j-1})\prod_{j=i+1}^n(\alpha^{i}-\alpha^j)}. 
\label{cpn}
\end{align}
Let us consider a melomorphic function  $f(z)=\frac{(C+z)^n}{z\prod_{j=1}^n(z-\alpha^j)}$ on $\bc\cup\{\infty\}$. $f(z)$ has simple poles at $z=0,\infty, \alpha^i\;\; (i=1,\cdots,n)$ and sum of residues at these points equals zero. Therefore, we have 
\begin{align}
\frac{C^n}{\prod_{i=1}^n(-\alpha^i)}
+\sum_{i=1}^{n} \frac{(C+\alpha^{i})^n}{\alpha^{i}\prod_{j=1}^{i}(\alpha^{i}-\alpha^{j-1})\prod_{j=i+1}^n(\alpha^{i}-\alpha^j)}-1=0.
\label{cpnr}
\end{align}
By combining (\ref{cpn}) with (\ref{cpnr}), we obtain
\begin{align}
\int_{\mathbb{C}P^n}(c_1(H))^n=1.
\end{align}

\section{The Model in This Paper}
\subsection{Base Model (Topological Sigma Model (A-Model) with Half BRST-Symmetry)}
We introduce our base model (topological sigma model (A-model)).
Lagrangian of the model is given as follows \cite{ref5}.
\begin{align}
L= \int_{\mathbb{C} \mathrm{P^1}} dz d\bar{z} \bigl[ &\frac{t}{2} g_{i \bar{j} } ( \partial_z \phi^i \partial_{\bar z}\phi^{\bar j} +  \partial_{\bar z} \phi^i \partial_z \phi^{\bar j}) +  \sqrt{t} i g_{i \bar{j}} \psi^{\bar{j}}_z  D_{\bar z}  \chi^i +\sqrt{t} ig_{i \bar{j}} {\psi}^i_{\bar z} D_z  \chi^{\bar{j}} \notag \\& - R_{i \bar{j} k \bar{l}} {\psi}^i_{\bar z} \psi^{\bar{j}}_z \chi^{k} \chi^{\bar l}\bigr].
\label{topA}
\end{align}
Fields in the Lagrangian and covariant derivatives are the same as the ones introduced in Section 1.

Original BRST-symmetry of this model is given in \cite{ref5}.
\begin{align}
\delta{\phi}^i  &= i \alpha \chi^i, & 
\delta \phi^{\bar i} &=i \alpha \chi^{\bar i}, & \notag \\
\delta \psi^{\bar{i}}_z &=-\sqrt{t}\alpha \partial_z \phi^{\bar i}-i \alpha \Gamma^{\bar i}_{\bar {\mu} \bar{\nu}}\chi^{\bar{\mu}} \psi^{\bar{\nu}}_z, & \notag \\
\delta{\psi}^i_{\bar z} &=-\sqrt{t }\alpha \partial_{\bar z} \phi^i-i \alpha \Gamma^i_{\mu \nu} \chi^\mu \psi^\nu_{\bar z},&
\delta \chi^i  &=\delta \chi^{\bar i} =0, &
\end{align}
where $\alpha$ is a fermionic parameter.
In order to include potential terms induced from a  holomorphic tangent vector field $K$, we have to change the BRST-symmetry 
in the following way (we observed that this change is inevitable to extend BRST-symmetry to the Lagrangian with potential terms). 
\begin{align}
\delta \phi^{\bar i} &=i \bar{\alpha} \chi^{\bar i}, &
\delta \psi^{\bar{i}}_z &=-i \bar{\alpha} \Gamma^{\bar i}_{\bar {\mu} \bar{\nu}}\chi^{\bar{\mu}} \psi^{\bar{\nu}}_z, & \notag \\
\delta{\psi}^i_{\bar z} &=- \sqrt{t } \bar{\alpha} \partial_{\bar z} \phi^i,  &
\delta{\phi}^i  &=\delta \chi^i  =\delta \chi^{\bar i} =0, 
\label{a1}
\end{align}
where $\bar \alpha$ is also a fermionic parameter. In the next subsection, we prove that the Lagrangian (\ref{topA})
remains invariant under this new BRST-symmetry. \textcolor{red}{}We can confirm that this transformation is the BRST-transformation in the following way.
 Let $Q$ be the generator of this transformation defined via the relation $\delta X =: i\alpha \{Q,X\}$ 
($X$ is a field that appears in the theory).  \textcolor{red}{}If $Q$ has  nilpotency, i.e. $ \{ Q,\{ Q,X \} \}=0$, this transformation is called  the BRST-transformation. 
The most non-trivial part comes from deriving $\{ Q,\{ Q,\psi^{\bar i}_{z} \} \}=0$. 
In this case, we have $\{ Q,\psi^{\bar i}_{ z} \} =- \Gamma^{\bar i }_{\bar \mu  \bar \nu } \chi^{\bar \mu} \psi^{\bar \nu}_{\bar z}$. 
By using the following relation that holds for the curvature tensor of  K$\ddot a$hler manifold,
\begin{align*}
R^{\bar i}_{ \bar \nu \bar l \bar \mu } = \partial _{ \bar l }\Gamma^{\bar i }_{\bar \mu  \bar \nu } - \Gamma^{\bar i }_{\bar \mu  \bar \alpha }\Gamma^{\bar \alpha }_{\bar l  \bar \nu } -  \partial _{ \bar \mu}\Gamma^{\bar i }_{\bar l  \bar \nu } + \Gamma^{\bar i }_{\bar l  \bar \alpha }\Gamma^{\bar \alpha }_{\bar \mu  \bar \nu } =0 \\
\Rightarrow  \partial _{ \bar l }\Gamma^{\bar i }_{\bar \mu  \bar \nu } - \Gamma^{\bar i }_{\bar \mu  \bar \alpha }\Gamma^{\bar \alpha }_{\bar l  \bar \nu } =\partial _{ \bar \mu}\Gamma^{\bar i }_{\bar l  \bar \nu } - \Gamma^{\bar i }_{\bar l  \bar \alpha }\Gamma^{\bar \alpha }_{\bar \mu  \bar \nu },
\end{align*}
we can show $\delta \bigl( - \Gamma^{\bar i }_{\bar \mu  \bar \nu } \chi^{\bar \mu} \psi^{\bar \nu}_{\bar z} \bigr)  =0$.  
Hence $\{ Q,\{ Q, \psi^{\bar i}_{ z} \} \} =0$ holds.
Check for other fields is straightforward.

\subsection{Proof of $\delta L=0$}
In this subsection, we check invariance of the Lagrangian in (\ref{topA}) under the BRST-transformation given in (\ref{a1})
, i.e., the equality $\delta L=0$.
We first evaluate variation of $g_{i \bar{j} } \partial_z \phi^i \partial_{\bar z}\phi^{\bar j}$.
\begin{equation}
\begin{split}
\delta \bigl( tg_{i \bar{j} } \partial_z \phi^i \partial_{\bar z}\phi^{\bar j} \bigr) 
&=t\partial_{\bar l} ( g_{i \bar{j} } ) \delta \phi^{\bar l}  \partial_z \phi^i \partial_{\bar z} \phi^{\bar j} +tg_{i \bar{j} } \partial_z \phi^i \partial_{\bar z} ( \delta \phi^{\bar j} )  \\
&=i t\bar \alpha \bigl(  g_{i \bar{\lambda} }\Gamma^{\bar \lambda}_{\bar{j} \bar{l}} \partial_z \phi^i \partial_{\bar z}\phi^{\bar j} \chi^{\bar l} +g_{i \bar{j} } \partial_z \phi^i \partial_{\bar z} \chi^{\bar j} \bigr) \\
&=i t\bar \alpha  g_{i \bar{j} } \partial_z \phi^i  \bigl( \partial_{\bar z}\chi^{\bar j} + \Gamma^{\bar j}_{\bar{\mu} \bar{\nu}}  \partial_{\bar z}\phi^{\bar \mu} \chi^{\bar \nu} \bigr) \\
&=i t\bar \alpha  g_{i \bar{j} } \partial_z \phi^i D_{\bar z} \chi^{\bar j}.
\end{split}
\end{equation}
By integration by parts, we obtain the following (we neglect total differential).
\begin{align}
 &\int_\Sigma dz d\bar{z}\delta \bigl(\frac{1}{2}  tg_{i \bar{j} } \partial_z \phi^i \partial_{\bar z}\phi^{\bar j} \bigr) \\
&= it \bar \alpha \int_\Sigma dz d\bar{z}\frac{1}{2}  \bigl( g_{i \bar{j} } \partial_z \phi^i  \partial_{\bar z}\chi^{\bar j} + g_{i \bar{j} } \partial_z \phi^i \Gamma^{\bar j}_{\bar{\mu} \bar{l}}  \partial_{\bar z}\phi^{\bar \mu} \chi^{\bar l} \bigr) \notag \\
&= i t\bar \alpha \int_\Sigma dz d\bar{z}\frac{1}{2}  \bigl(  -\partial_l g_{i \bar{j} } \partial_{\bar z} \phi^l \partial_z \phi^i \chi^{\bar j} -\partial_{\bar l} g_{i \bar{j} } \partial_{\bar z} \phi^{\bar l} \partial_z \phi^i \chi^{\bar j}  \notag \\
& - g_{i \bar{j} }  \partial_{\bar z} \partial_z \phi^i \chi^{\bar j}  + \partial_{\bar \mu} g_{i \bar{l} } \partial_{\bar z} \phi^{\bar \mu} \partial_z \phi^i \chi^{\bar l}  \bigr)  \notag \\
&= i t\bar \alpha \int_\Sigma dz d\bar{z} \frac{1}{2} \bigl[  g_{i \bar{j} } \partial_{\bar z} \phi^{ i } \bigl( \partial_z \chi^{\bar j } + \Gamma^{\bar j  }_{\bar{ \mu } \bar{ \nu }} \partial_z \phi^{\bar \mu }  \chi^{\bar \nu } \bigr) \bigr] \notag \\
&= i t\bar \alpha \int_\Sigma dz d\bar{z} \frac{1}{2} \bigl(  g_{i \bar{j} } \partial_{\bar z} \phi^{ i }D_z \chi^{\bar j }  \bigr).
\end{align}
Variation of other terms are evaluated as follows.
\begin{align}
\delta \bigl(\frac{1}{2} tg_{i \bar{j} } \partial_z \phi^{ \bar j } \partial_{\bar z}\phi^{ i } \bigr) 
&=  i t\bar \alpha \frac{1}{2}  \bigl(  g_{i \bar{ j } }  \Gamma^{\bar j }_{\bar{ l } \bar{ \mu }} \chi^{\bar l } \partial_z \phi^{\bar \mu } \partial_{\bar z} \phi^{ i }+  tg_{i \bar{j} }  \partial_{\bar z}\phi^{ i }  \partial_z \chi^{\bar j }  \bigr) \notag \\
&= i t\bar \alpha\frac{1}{2}   g_{i \bar{j} } \partial_{\bar z} \phi^{ i }D_z \chi^{\bar j }.  \\
\delta \bigl( \sqrt{t} ig_{i \bar{j}} \psi^{\bar{j}}_z  D_{\bar z}  \chi^i  \bigr) 
&=- \sqrt{t} \bar \alpha R_{ i \bar { j }  k \bar {l}  } \partial_{\bar z}\phi^{ i } \psi^{\bar{j}}_z  \chi^k  \chi ^{\bar { l }}.  \\
\delta \bigl(  \sqrt{t}i g_{i \bar{j}} {\psi}^i_{\bar z} D_z  \chi^{\bar{j}} \bigr) 
&= \sqrt{t}  \bar \alpha g_{i \bar{j}} ( \partial_{ \bar l } \Gamma^{\bar j }_{\bar{ \mu }  \bar{ \nu } }  +  \Gamma^{\bar j }_{\bar{ \beta } \bar{ l }} \Gamma^{\bar \beta }_{\bar{ \mu } \bar{ \nu }}  ) {\psi}^i_{\bar z}  \chi^{\bar l } \partial_{ z} \phi^{ \bar \mu } \chi ^{\bar { \nu }} - ti \bar \alpha g_{i \bar{ j } } \partial_{\bar z}\phi^{ i } D_z  \chi^{\bar{j}}.
\end{align}
By using the following  equality that holds for the  curvature of a K$\ddot{\mbox a}$hler manifolds:
\begin{equation}
R^{ \bar j }_{ \bar \nu \bar l \bar \mu } = \partial_{ \bar l } \Gamma^{\bar j }_{\bar{ \nu }  \bar{ \mu } } - \partial_{ \bar \mu } \Gamma^{\bar j }_{\bar{ l }  \bar{ \nu } } + \Gamma^{\bar j }_{\bar{ \beta } \bar{ l }} \Gamma^{\bar \beta }_{\bar{ \mu } \bar{ \nu }} -\Gamma^{\bar j }_{\bar{ \beta } \bar{\mu }} \Gamma^{\bar \beta }_{\bar{ \nu } \bar{ l }}  = 0.
\end{equation}
and K$\ddot{\mbox a}$hler condition $\Gamma^{\bar \beta }_{\bar{ \mu } \bar{ \nu }} = \Gamma^{\bar \beta }_{\bar{ \nu } \bar{ \mu }}$, we obtain
\begin{align}
&\delta \bigl( \sqrt{t} i g_{i \bar{j}} {\psi}^i_{\bar z} D_z  \chi^{\bar{j}} \bigr) =- ti \bar \alpha g_{i \bar{ j } } \partial_{\bar z}\phi^{ i } D_z \chi^{\bar{j}}. 
\end{align}
Next, we use the following formula of covariant derivative of the curvature,
\begin{equation}
\begin{split}
\triangledown_{\bar{ \lambda } } R_{i \bar{j} k \bar{l}} =\partial _{\bar { \lambda }} R_{i \bar{j} k \bar{l}} -R_{i \bar{ \beta } k \bar{l}}  \Gamma^{\bar \beta  }_{ \bar{ \lambda  } \bar{ j } }  - R_{i \bar{ j } k \bar{ \beta }}  \Gamma^{\bar \beta  }_{ \bar{ \lambda } \bar{ l } }, 
\end{split}
\end{equation}
and Bianchi's identity,
\begin{equation}
\begin{split}
\triangledown_{\bar{ \lambda } } R^{i }_{ \bar{j} k \bar{l}} =\triangledown_{\bar{ l } } R^{ i }_{ \bar{j} k \bar{ \lambda }}.
\end{split}
\end{equation}
Then we obtain
\begin{equation}
\begin{split}
\delta \bigl( -  R_{i \bar{j} k \bar{l}} {\psi}^i_{\bar z} \psi^{\bar{j}}_z \chi^{k} \chi^{\bar l}  \bigr) 
&= -  i\bar \alpha \triangledown_{\bar{ \lambda } }  R_{i \bar{j} k \bar{l}}  {\psi}^i_{\bar z} \psi^{\bar{j}}_z \chi^{k} \chi^{\bar l} \chi^{ \bar \lambda }  +\sqrt{t} \bar \alpha  R_{i \bar{j} k \bar{l}} \partial_{ \bar z } \phi^{ i }  \psi^{\bar{j}}_z \chi^{k} \chi^{\bar l} \\
&=\sqrt{t} \bar \alpha  R_{i \bar{j} k \bar{l}} \partial_{ \bar z } \phi^{ i }  \psi^{\bar{j}}_z \chi^{k} \chi^{\bar l}. 
\end{split}
\end{equation}
As a result, all the variations cancel each other.  
Therefore, we have shown the equality: $\delta L=0 \Leftrightarrow \{Q,L\}=0$.

\subsection{BRST-Closed Observables of the Base Model}
In this subsection, we consider BRST-closed observable of this model, i.e.,  observable $\mathcal{O}$ that satisfies $\{Q,\mathcal{O}\}=0 $. Here we restrict observables that are obtained from differential forms on $M$. 
Let $W$ be  a $(p,q)$-form on $M$:
\begin{equation}
W=\frac{1}{p!q!} W_{ i_1 i_2  \cdots  i_p \bar{ j_1} \bar{ j }_2  \cdots \bar{j_q} } (z^1,\cdots z^m) dz^{i_1} dz^{i_2}  \cdots dz^{i_p} d{\bar z}^{\bar { j_1}} d{\bar z}^{\bar { j_2}} \cdots d{\bar z}^{\bar { j_q}},
\end{equation}
then we consider the following observable $\mathcal{O}_W$:
\begin{equation}
\mathcal{O}_W=\frac{1}{p!q!} W_{ i_1 i_2  \cdots  i_p \bar{ j_1} \bar{ j }_2  \cdots \bar{j_q} } (\phi)\chi^{i_1} \chi^{i_2}  \cdots \chi^{i_p} \chi^{\bar { j_1}} \chi^{\bar { j_2}} \cdots \chi^{\bar { j_q}}.
\end{equation}
Variation of $\mathcal{O}_W$ under the BRST-transformation:
\begin{align*}
\{Q,\phi^i\}=0,\{Q,\phi^{\bar i}\}=\chi^{\bar i},\{Q,\chi^i\}=0,\{Q,\chi^{\bar i}\}=0,
\end{align*}
is given by
\begin{align}
\delta \mathcal{O}_W &= \frac{1}{ p! q! } \partial_{\bar l }W_{ i_1 i_2  \cdots  i_p \bar{ j_1} \bar{ j }_2  \cdots \bar{j_q} } \delta \phi^{\bar l } \chi^{i_1} \chi^{i_2}  \cdots \chi^{i_p} \chi^{\bar { j_1}} \chi^{\bar { j_2}} \cdots \chi^{\bar { j_q}}   \notag  \\
&= i \bar \alpha \frac{1}{ p! q! } \partial_{\bar l }W_{ i_1 i_2  \cdots  i_p \bar{ j_1} \bar{ j }_2  \cdots \bar{j_q} } \chi^{\bar l } \chi^{i_1} \chi^{i_2}  \cdots \chi^{i_p} \chi^{\bar { j_1}} \chi^{\bar { j_2}} \cdots \chi^{\bar { j_q}}.   
\end{align}
The above result is summarized 
as follows.
\begin{equation}
\delta \mathcal{O}_W = i \bar \alpha \bar \partial \mathcal{O}_W = i \bar \alpha \{Q,\mathcal{O}_W\}.
\end{equation}
$\{Q,~\}$ is represented as follows.
\begin{equation*}
\{Q,\mathcal{O}_W\}=\mathcal{O}_{\bar \partial W}.
\end{equation*}
Therefore, a BRST-closed observable $\mathcal{O}_{W}$ is obtained from a differential form $W$
that satisfies $\bar{\partial} W=0$.
By standard discussion of topological field theory, correlation function of BRST-closed observables with insertion of 
a observable of type $\{Q,\mathcal{O}_W\}=\mathcal{O}_{\bar \partial W}$ automatically vanishes. 
Hence, physical observables of the base model correspond to elements of Dolbeault cohomology. 
Let us recall Dolbeault's theorem and Hodge's decomposition theorem.
\begin{theorem} (Dolbeault's theorem)
\begin{equation}
H^q(M,\wedge^{p}T^{\prime *}M) \simeq H^{p,q}_{\bar \partial}(M).
\end{equation}
\end{theorem}
\begin{theorem} (Hodge's decomposition thorem)\cite{ref14}
\begin{align}
H^r(M, \mathbb{C})\simeq \displaystyle{\bigoplus_{p+q=r}} H^{p,q}(M). \\
H^{p,q}(M) \simeq  H^{p,q}_{\bar \partial}(M) \simeq H^q(M,\wedge^{p}T^{\prime *}M) . 
\end{align}
\end{theorem}
If $W$ is a Chern class of a holomorphic vector bundle of $M$, it is given as a closed $(i,i)$-form on $M$.
Therefore $\d W=\bar{\d} W=0$ follows from $dW=(\d+\bar{\d})W=0$ and ${\cal O}_{W}$ is a BRST-closed 
observable of the base model.

\subsection{ Potential Terms Induced from Holomorphic Tangent Vector Field}
In this subsection, we include potential terms induced from the holomorphic tangent vector field $K$. The potential terms are given as follows (we use  a parameter $\beta$ that equals $2\pi i$ for brevity).
\begin{equation}
V= \int_\Sigma dz d\bar{z}\bigl[  t s^2 \beta g_{i \bar{j}} K^i \bar{K}^{\bar{j}} + t sg_{i \bar{j}} \triangledown_{\bar{\mu}} \bar{K}^{\bar{j}} \chi^{\bar {\mu}} \chi^i + s\beta g_{i \bar{j}} \triangledown_{\mu} K^{i}  \psi^{\mu}_{\bar{z}}  \psi^{\bar{j}}_z  \bigr].
\end{equation}
$V$ contains a parameter $s\in \mathbb{R}$ that controls scale of the vector field $K$. 
We can extend the BRST-transformation to the new Lagrangian $L+V$ as follows.
\begin{align}
\delta \phi^{\bar i} &=i \bar{\alpha} \chi^{\bar i}, &  \delta \psi^{\bar{i}}_z &=-i \bar{\alpha} \Gamma^{\bar i}_{\bar {\mu} \bar{\nu}}\chi^{\bar{\mu}} \psi^{\bar{\nu}}_z,  &  \delta{\psi}^i_{\bar z} &=- 2\sqrt{t}\bar{\alpha} \partial_{\bar z} \phi^i,  &\\
\delta \chi^i &= i s\bar \alpha \beta K^i,  & \delta{\phi}^i & =\delta \chi^{\bar i} =0.   
\label{a2}
\end{align}
Let $Q$ be \textcolor{red}{}the generator of this transformation whose action is defined via the relation  $\delta X =: i\alpha \{Q,X\}$ ($X$ is a field that appears in the theory).  We check nilpotency of this generator $ \{ Q,\{ Q,X \} \}=0$. 
Non-trivial parts caused by appearance of $K$ in (\ref{a2}) is given as follows.
\begin{align*}
\delta \phi^{ \bar i} &= i \bar \alpha \{ Q,\phi^{ \bar i} \} \Rightarrow \{ Q,\phi^{ \bar i} \} =\chi^{\bar i},  & \\ 
\delta \chi^{\bar i} &= i \bar \alpha \{ Q, \{ Q,\phi^{ \bar i} \} \} \Rightarrow \{ Q, \{ Q,\phi^{ \bar i} \} \}=0, &   \\
  \delta \chi^i &= i \bar \alpha \{ Q, \chi^i \}  \Rightarrow \{ Q, \chi^i \} =\beta  sK^i, &\\
\delta K^i &=i \bar \alpha \{ Q,\{ Q, \chi^i \} \} \Rightarrow \{ Q,\{ Q, \chi^i \} \} =0.&  
\end{align*}
Hence the relation $ \{ Q,\{ Q,X \} \}=0$ also holds in this case.

\subsection{Proof of $\delta(L+V)=0$}
In this subsection, we check invariance of the Lagrangian $L+V$ under (\ref{a2}), i.e., the relation $\delta(L+V)=0$.
Let us recall some properties of covariant derivatives of K.
\begin{align}
&\triangledown_{\bar{\mu}} K^{ j } =0, \quad\quad  \triangledown_{\mu} \bar{K}^{\bar{j}} =0, \\
&\partial_{\bar l } \bigl(g_{i \bar{j}} \triangledown_{\bar{\mu}} \bar{K}^{\bar{j}} \bigr) = g_{i \bar{j}} \triangledown_{\bar{ l }} \triangledown_{\bar{\mu} } \bar K^{\bar{ j } }  +g_{i \bar{ j } }\Gamma^{\bar \alpha}_{\bar l  \bar \mu } \triangledown_{\bar{\alpha} } \bar K^{\bar{ j } },   \\& \partial_{\bar l } \bigl(g_{i \bar{j}} \triangledown_\mu K^i \bigr) = g_{i \bar{ \lambda } }\Gamma^{\bar \lambda }_{\bar j  \bar l }\triangledown_\mu K^i  +  R_{ i \bar j \mu \bar l } K^i.
\end{align}
We use the above formulas and standard property of K$\ddot{\mbox a}$hler metric. 
Then additional terms that appear in checking $\delta ( L + V )$ are given as follows.
\begin{align}
\delta \bigl(  \sqrt{t} i g_{i \bar{j}} \psi^{\bar{j}}_z  D_{\bar z}  \chi^i   \bigr) 
  &=\sqrt {t}s\bar \alpha \beta g_{i \bar{j}} \psi^{\bar{j}}_z\triangledown_{ l } K^{ i }   \partial_{ \bar z } \phi^l, \\
\delta \bigl( -R_{i \bar{j} k \bar{l}} {\psi}^i_{\bar z} \psi^{\bar{j}}_z \chi^{k} \chi^{\bar l}  \bigr) 
&=-s\bar \alpha \beta i R_{i \bar{j} k \bar{l}} {\psi}^i_{\bar z} \psi^{\bar{j}}_z K^{k} \chi^{\bar l}, \\
\delta \bigl( ts^2\beta g_{i \bar{j}} K^i \bar{K}^{\bar{j}}  \bigr)  
&=t s^2i \bar \alpha \beta  g_{i \bar{j}} K^i \triangledown_{\bar l } \bar{K}^{\bar j } \chi^{\bar l },\\
\delta \bigl( tsg_{i \bar{j}} \triangledown_{\bar{\mu}} \bar{K}^{\bar{j}} \chi^{\bar {\mu}} \chi^i  \bigr) 
&=  - ti s^2\bar \alpha  \beta g_{i \bar{j}} \triangledown_{\bar{\mu}} \bar{K}^{\bar{j}}   K^i \chi^{\bar {\mu}}, \\
\delta \bigl( s\beta g_{i \bar{j}} \triangledown_{\mu} K^{i}  \psi^{\mu}_{\bar{z}}  \psi^{\bar{j}}_z  \bigr)
&= s \bar \alpha  \beta i R_{ i \bar j k \bar l }  \psi^{i}_{\bar{z}}     \psi^{\bar{j}}_z K^k \chi^{ \bar l }  -\sqrt{t} s\bar \alpha \beta  g_{i \bar{j}} \triangledown_{\mu} K^{i} \partial_{\bar z} \phi^\mu  \psi^{\bar{j}}_z.
\end{align}
From the above results, we can conclude that $\delta ( L + V )=0$ holds. From now on, we only consider the model 
given by $L+V$.

\subsection{BRST-Closed Observable of the Model}
In this subsection, we construct BRST-closed observable of the model with potential terms. 
In the same way as the previous discussions, we restrict our selves to observables obtained from a $(p,q)$-form $W$ on $M$. 
It is represented in the following form. 
\begin{equation}
{\cal O}_{W}=\frac{1}{p!q!} W_{ i_1 i_2  \cdots  i_p \bar{ j_1} \bar{ j }_2  \cdots \bar{j_q} } \chi^{i_1} \chi^{i_2}  \cdots \chi^{i_p} \chi^{\bar { j_1}} \chi^{\bar { j_2}} \cdots \chi^{\bar { j_q}}.
\end{equation}
Variation $\delta {\cal O}_{W}$.under the BRST-transformation is given by
\begin{align}
\delta {\cal O}_{W}&= \frac{1}{ p! q! } \partial_{\bar l }W_{ i_1 i_2  \cdots  i_p \bar{ j_1} \bar{ j }_2  \cdots \bar{j_q} } \delta \phi^{\bar l } \chi^{i_1} \chi^{i_2}  \cdots \chi^{i_p} \chi^{\bar { j_1}} \chi^{\bar { j_2}} \cdots \chi^{\bar { j_q}}   \notag  \\
&+  \frac{1}{ (p-1)!q!} W_{ l i_1 i_2  \cdots  i_{p-1}  \bar{ j_1} \bar{ j }_2  \cdots \bar{j_q} }  \delta \chi^{ l }  \chi^{i_1} \chi^{i_2}  \cdots \chi^{i_(p-1) } \chi^{\bar { j_1}} \chi^{\bar { j_2}} \cdots \chi^{\bar { j_q}}  \notag  \\
&= i \bar \alpha \bigl[\frac{1}{ p! q! } \partial_{\bar l }W_{ i_1 i_2  \cdots  i_p \bar{ j_1} \bar{ j }_2  \cdots \bar{j_q} } \chi^{\bar l } \chi^{i_1} \chi^{i_2}  \cdots \chi^{i_p} \chi^{\bar { j_1}} \chi^{\bar { j_2}} \cdots \chi^{\bar { j_q}}   \notag  \\
&+  \frac{s\beta }{ (p-1)!q!} W_{ l i_1 i_2  \cdots  i_{p-1}  \bar{ j_1} \bar{ j }_2  \cdots \bar{j_q} }  K^{ l }  \chi^{i_1} \chi^{i_2}  \cdots \chi^{i_{p-1} } \chi^{\bar { j_1}} \chi^{\bar { j_2}} \cdots \chi^{\bar { j_q}}  \bigr].
\end{align}
 Let $i(K)$ be the inner-product operator by $K$. Then, the above result is rewritten as follows.
\begin{equation}
\delta{\cal O}_{W}= i \bar \alpha {\cal O}_{(\bar \partial + \beta i(sK)) W }=: i \bar \alpha \{Q,{\cal O}_{W}\}.
\label{brst2}
\end{equation}
Hence  BRST-closed observable is obtained from a differential form $W$ that satisfies $(\bar \partial + \beta i(sK))W=0$. 
Note that this condition reduces to $\bar \partial W=0$ if $s = 0$. 

We comment on mathematical background to this condition.  In general, differential form $\omega$ on $M$ is graded by the following 
operators:
\begin{align*}
F_A \omega &=(p+q) \omega,& 
F_V \omega &=(q-p) \omega.&
\end{align*}
This says that $\omega$ is a $(p,q)$-form on $M$.
For the operator $\bar \partial + \beta i(sK)$, we adopt $F_V$ as the grading operator and consider the following 
vector space:
\begin{align*}
A^{(k)}=\displaystyle \bigoplus_{q-p=k} \Omega^{p,q} (M)
\end{align*}
where $k$ ranges from $-m$ to $m$. From the condition (\ref{brst2}), we can see that observables correspond to elements of cohomology of the complex $(A^{(k)}, \bar \partial +\beta i(sK))$. This complex is called Liu's complex. In \cite{ref15}, it is shown that cohomology of 
this complex is independent of $s\;\; (s\neq0)$. This property is closely related to Proposition 1.

\section{Derivation of the Bott Residue Formula}
\subsection{Overview}
In this subsection, we explain our strategy of deriving the Bott 
residue formula by using the topological sigma model with potential terms. 

 Since the Bott residue formula is a fixed point formula for integration of Chern classes of the holomorphic vector bundle $E$,
we consider observable that corresponds to wedge product of Chern classes in the $s\to0$ limit. 
We first construct observable ${\cal O}_{W}$ that satisfies $(\bar \partial +\beta i(sK))W=0$ and 
$\lim_{s\to 0}W=\underline{\varphi}(E)$.    
Let us recall Proposition 1 introduced in Subsection 1.2.
\begin{quotation}
\noindent
{\bf Proposition 1}\\
\\ 
Correlation functions of BRST-closed observables are  invariant under variation of $s$.
\\
\end{quotation}
Assuming its proposition, we evaluate the degree $0$ correlation function $\langle {\cal O}_{W}\rangle_{0}$
both in the $s \to 0$ limit and $s \to \infty$ limit.

As for the $s\to 0$ limit, the observable becomes ${\cal O}_{\underline{\varphi}(E)}$ and we can use standard weak coupling limit. 
Moreover, the potential terms vanish in this limit.
Then we expand each field around the solution of the classical equations of motion ($\phi=\phi_{0}$ (constant map), $\chi=\chi_{0}$ (constant solution), $\psi=0$) and perform Gaussian integration of oscillation modes. 
We show that contributions from Gaussian integral is trivial. 
Hence $\langle {\cal O}_{W}\rangle_{0}$ turns out to be classical integration of $(m,m)$-differential form
$\underline{\varphi}(E)$ on $M$, i.e., the l.h.s of (\ref{goal}). 

In the $s \to \infty$ limit, the path integral is localized around neighborhood of zero set $\{M_{\alpha}\}$ of the holomorphic vector 
field $K$ where the condition $\delta \chi^{i}=is\alpha\beta K^{i}=0$ holds. We also expand each field around the solution of the classical equations of motion ($\phi=\phi_{0}\in M_{\alpha}$  
(constant map), $\chi=\chi_{0}$ (constant solution), $\psi=0$).
In this case, we carefully discuss integration measure of oscillation modes by using  eigenvalue decomposition by Laplacian for 
differential forms on $\mathbb{C}P^{1}$.  
Since contributions from oscillation modes do not affect correlation functions,  contribution from a connected component $M_{\alpha}$ turns out to be integration of 
differential form on $M_{\alpha}$ given in the r.h.s of (\ref{goal}). 
Summing up all the connected components, $\langle {\cal O}_{W}\rangle_{0}$ becomes the r.h.s of (\ref{goal}). 

We can equate these two results by using Proposition 1 and obtain the Bott residue formula. 

\subsection{The BRST-Closed  Observable Used for Derivation }
In this subsection, we construct the observable ${\cal O}_{W}$ that satisfies $\lim_{s\to 0}W=\underline{\varphi}(E)$ and $(\bar \partial + \beta  i(K))W=0$. \textcolor{red}{} We use the notation in subsection 2.1. If we rescale $K$ into $sK$, $\theta(K)$ is also rescaled to $s\theta(K)$. In general, $\Lambda$ is also rescaled $s \Lambda$. Let $\{e_a\}$ be the local holomorphic frame of $E$. 
Then, we define $L=(L^{b}_{a}) : \Gamma (E) \to \Gamma (E) $ by
\begin{align}
L(s):=s\Lambda (s)-i(sK) \tilde \nabla (s)
\end{align}
($s \in  \Gamma (E)$), where we rescale $K$ into $sK$. For the local holomorphic frame,
\begin{align}
 L^b_a e_b  :=s\Lambda^b_a  e_b - s\Theta^b_{ a k } K^k e_b.
\end{align}
Let us note that the following relations hold. 
\begin{align}
&\bar \partial  i(sK) \tilde{\nabla}e_a  = s\partial_{ \bar l }(\Theta^b_{ a k } ) K^k d\overline{z^{l}} e_b = -i(sK)(F^b_{ a k \bar l }dz^{k}\wedge d\overline{z^{l}} e_b ), & 
\notag \\
&\bar \partial ( L^b_a e_b )  = \bar \partial (s \Lambda^b_a e_b - s\Theta^b_{ a k } K^k e_b ) =-s\partial_{ \bar l }(\Theta^b_{ a k } ) K^k  d\overline{z^{l}} e_b = i(sK)(F^b_{ a k \bar l }dz^{k}\wedge d\overline{z^{l}}e_b ),  & \notag \\
&\bar \partial ( F^a_{ b k \bar l } dz^{k}\wedge d\overline{z^{l}} e^b )  =0, \notag \\
& i(sK)( L^a_b e^b ) =0.
\end{align}
By using the above relations, we obtain ($\beta=2\pi i$),
\begin{align}
(\bar \partial + \beta i(sK) ) (L^b_a e_b+ \frac{i}{2\pi }F^b_a e_b)=\bar \partial ( L^b_a e_b )-i(sK)(F^b_{ a k \bar l } dz^{k}\wedge d\overline{z^{l}} e_b )=0.
\label{kuwakey}
\end{align}
If we define matrix valued form:
\ba
A=(A^{a}_{b}),\; A=  L+ \frac{i}{2\pi} F,
\ea 
(\ref{kuwakey}) says $(\bar \partial + \beta  i(K)) A=0$. 
Therefore $(\bar \partial + \beta  i(K))\mathrm{tr}(A^m)=0$ holds for arbitrary positive integer $m$.
Let $U$ be a linear automorphism of a complex vector space $V$ with $\dim_{\mathbb{C}}=\mathrm{rank}(E)=q$.
It is well-known that $\varphi(U)$ defined in Subsection 2.1 can be represented in the following way.  
\begin{equation}
 \varphi(U)=\sum_{m_{i}\geq 0,\;\sum_{i=1}^{l}m_{i}=m}\alpha_{m_{1}m_{2}\cdots m_{i}}
\mathrm{tr}(U^{m_1})\mathrm{t}r(U^{m_2}) \cdots \mathrm{tr}(U^{m_l }),
\end{equation}
where $m$ is the complex dimension of $M$.
Therefore,  $\varphi(A)=\varphi(L+ \frac{i}{2\pi} F)$ is annihilated by the operator $\bar \partial + \beta i(sK)$.
Obviously, $\varphi(L+ \frac{i}{2\pi} F)$ reduces to $\varphi(\frac{i}{2\pi} F)=\underline{\varphi}(E)$  under the $s\to 0$ limit.
In this way, we have constructed the operator ${\cal O}_{\varphi(L+ \frac{i}{2\pi} F)}$ that is used in derivation of the Bott residue 
formula. From now on, we simply denote it by $\varphi$ for brevity.

\subsection{Degree $0$ Correlation Function and Integral Measure in the $s \to 0$ Limit}
From now on, we consider the degree $0$ correlation function $<\varphi>_{0}$. 
First, we rewrite the Lagrangian in the following form.
\begin{align}
L+V&= \frac{t}{2} \int_\Sigma \phi^* (\omega) +L'+V',\\
L'+V'&:=\int_{\mathbb{C} \mathrm{P^1}} dz d\bar{z} \bigl[t g_{i \bar{j} }  \partial_{\bar z} \phi^i \partial_z \phi^{\bar j} +  \sqrt{t}i g_{i \bar{j}} \psi^{\bar{j}}_z  D_{\bar z}  \chi^i +\sqrt{t}ig_{i \bar{j}} {\psi}^i_{\bar z} D_z  \chi^{\bar{j}} \notag \\&- R_{i \bar{j} k \bar{l}} {\psi}^i_{\bar z} \psi^{\bar{j}}_z \chi^{k} \chi^{\bar l} +ts^2\beta g_{i \bar{j}} K^i \bar{K}^{\bar{j}} + t sg_{i \bar{j}} \triangledown_{\bar{\mu}} \bar{K}^{\bar{j}} \chi^{\bar {\mu}} \chi^i + s\beta g_{i \bar{j}} \triangledown_{\mu} K^{i}  \psi^{\mu}_{\bar{z}}  \psi^{\bar{j}}_z \bigr]
\end{align}
where $\omega=g_{i\bar{j}}dz^{i}\wedge d\overline{z^{j}}$ is the K$\ddot{\mbox a}$hler form of $M$.
 $\int_\Sigma \phi^* (\omega)$ is a topological term that gives mapping degree of $\phi: \mathbb{C}P^{1}\to M$. Since we focus on 
the degree $0$ correlation function, we use the Lagrangian $L'+V'$ instead of $L+V$. 
By using  Proposition 1,
we obtain the following equality. 
\begin{equation}
\displaystyle \lim_{s \to 0} <\varphi>_{0}=\displaystyle \lim_{s \to \infty} <\varphi>_{0}.
\end{equation}
In this subsection, we focus on the left hand side. 
In the $s \to 0$ limit, the Lagrangian $L+V$ becomes Lagrangian of the usual topological sigma model (A-model). 
Moreover,  $\varphi$ turns into ${\cal O}_{\underline{\varphi}(E)}$, which is also a standard BRST-closed observable of 
the A-model.  
Then we can apply standard result of the weak coupling limit $t \to \infty$ \cite{ref5}.
It says that the path integral reduces to Gaussian integration around the constant map $\phi(z,\bar{z})=\phi_{0}\;(\in M)$. 
Then the correlation function becomes,
\begin{align*}
\displaystyle \lim_{s \to 0} <\varphi>_{0}
=\int_M d\phi_0 d\chi_0{\cal O}_{\underline{\varphi}(E)}=\int_{M}\underline{\varphi}(E)=:\underline{\varphi}(E)[M],
\end{align*}
where $d\phi_0 d\chi_0$ is the measure for integration of position of $\phi_{0}\in M$ and the corresponding zero-mode of $\chi$, 
that  can be interpreted as integration of $(m,m)$-form on $M$.

\subsection{Integration Measure for  the Degree $0$ Correlation Function in the $s \to \infty$ Limit}
\label{measure}
We discuss integration measure in evaluating the correlation function in the $s \to \infty $ limit with fixed $t$. 
Since we are considering degree $0$ correlation function, the map $\phi$ is homotopic to a constant map 
$\phi(z,\bar{z})=\phi_{0}\;(\in M)$. Therefore, we expand the fields $\phi$, $\chi$ and $\psi$ around the
constant map $\phi_{0}$. Then $\chi$ (resp. $\psi$) becomes section of $\phi_{0}^{-1}(T^{\prime}M)$ 
(resp. $\phi_{0}^{-1}(T^{\prime}M)\otimes \overline{T^{\prime *}}\mathbb{C}\mathrm{P^1}$) and its complex 
conjugate. But $\phi_{0}^{-1}(T^{\prime}M)$ is isomorphic to trivial bundle $\mathbb{C}\mathrm{P^1}\times\mathbb{C}^m$.
Hence we can simply regard $\chi^{i}, \chi^{\bar{i}}$ (resp. $\psi^{i}_{\bar{z}}, \psi^{\bar{i}}_{z}$) as $(0,0)$-form (resp. $(0,1)$-form,
$(1,0)$-form) on $\mathbb{C}\mathrm{P^1}$    
By using standard K$\ddot{\mbox a}$hler metric of $\mathbb{C}\mathrm{P^1}$, we can apply eigenvalue decomposition by Laplacian for differential forms on  $\mathbb{C}\mathrm{P^1}$ to the expansion.
The Laplacian is represented as follows ($\dag $ means adjoint defined by Hodge operator of $\mathbb{C}\mathrm{P^1}$).
\begin{align*}
\triangle&:=dd^\dag +d^\dag d, & \triangle_\partial&:=\partial \partial^\dag +\partial^\dag \partial,& \triangle_{\bar \partial}:=\bar \partial \bar \partial^\dag +\bar \partial^\dag \bar \partial,&
\end{align*}
\begin{align*}
\triangle= 2\triangle_\partial=2\triangle_{\bar \partial}.
\end{align*}
Let $\triangle^{(p,q)}$ be restriction to $\Omega^{(p,q)}(\mathbb{C}\mathrm{P^1})$. Vector space of $(p,q)$-forms with 
zero eigenvalue is known as $H^{p,q}(\mathbb{C}\mathrm{P^1})$: the vector space of $(p,q)$ harmonic forms.  
The following result is well-known.
\ba
\dim_{\mathbb{C}}(H^{0,0}(\mathbb{C}\mathrm{P^1}))=1,\;\; 
\dim_{\mathbb{C}}(H^{1,0}(\mathbb{C}\mathrm{P^1}))=\dim_{\mathbb{C}}(H^{0,1}(\mathbb{C}\mathrm{P^1}))=0. 
\label{hcp1}
\ea
Let $\{ E_{n}\;|\;n>0\}$ be set of positive eigenvalues of $\frac{1}{2}\triangle^{(0,0)}$ ordered as follows.
\ba
0<E_{1}\leq E_{2}\leq E_{3}\leq \cdots. 
\ea
Then we denote by $f_{n}(z,\bar{z})$ the $(0,0)$-form that satisfy
\ba
\frac{1}{2}\triangle^{(0,0)}f_{n}(z,\bar{z})=E_{n}f_{n}(z,\bar{z}).
\ea 
\begin{lem}
Sets of positive eigenvalues of $\frac{1}{2}\triangle^{(1,0)}$ and $\frac{1}{2}\triangle^{(0,1)}$ are both given by $\{ E_{n}\;|\;n>0\}$
and $(1,0)$ and $(0,1)$ forms with eigenvalue $E_{n}$ are given by $\d f_{n}(z,\bar{z})$ and $\bar{\d} f_{n}(z,\bar{z})$ 
respectively. 
\end{lem}  
{\it Proof)}\\
Since $\frac{1}{2}\triangle$ equals $(\d{\d}^{\dag}+{\d}^{\dag}\d)$, we obtain
\ba
\frac{1}{2}\triangle\d f_{n}(z,\bar{z})&=&(\d{\d}^{\dag}+{\d}^{\dag}\d)\d f_{n}(z,\bar{z})\no\\
                              &=&\d{\d}^{\dag}\d f_{n}(z,\bar{z})\no\\
                              &=&\d(\d{\d}^{\dag}+{\d}^{\dag}\d)f_{n}(z,\bar{z})\no\\
                              &=&\frac{1}{2}\d\triangle  f_{n}(z,\bar{z})\no\\
                              &=&E_{n}\d f_{n}(z,\bar{z}). 
\ea 
Hence $\d f_{n}(z,\bar{z})$ is $(1,0)$-form with eigenvalue $E_{n}$. On the contrary, let $\omega$ be $(1,0)$ form with 
eigenvalue $E$. Then $(0,0)$ form $\d^{\dag}\omega$ satisfy
\ba
\frac{1}{2}\triangle\d^{\dag} \omega&=&(\d{\d}^{\dag}+{\d}^{\dag}\d)\d^{\dag} \omega\no\\
                              &=&{\d}^{\dag}\d \d^{\dag}\omega\no\\
                              &=&\d^{\dag}(\d{\d}^{\dag}+{\d}^{\dag}\d)\omega\no\\
                              &=&\frac{1}{2}\d^{\dag}\triangle \omega\no\\
                              &=&E\d^{\dag}\omega. 
\ea 
Therefore, $\d^{\dag}\omega$ must coincide some $f_{n}(z,\bar{z})$. This completes proof for $\triangle^{(1,0)}$.
Proof for $\frac{1}{2}\triangle^{(0,1)}$ goes in the same way by using the equality $\frac{1}{2}\triangle=(\bar{\d}{\bar{\d}}^{\dag}+\bar{\d}^{\dag}\bar{\d})$.
$\Box$   

Variation $\delta\phi$ from the constant map $\phi_{0}$ can also be regard as $(0,0)$ form on $\mathbb{C}\mathrm{P^1}$. 
Combining 
\begin{align}
\phi^i=\phi^i_0 +\displaystyle \sum_{n>0} \phi'^i_n f_n(z,\bar z),\quad
\phi^{\bar i}=\phi^{\bar i}_0 +\displaystyle \sum_{n>0} \phi'^{\bar i}_n \bar f_n(z,\bar z),
\label{ex1}
\end{align}
\begin{align}
\chi^i=\chi^i_0 +\displaystyle \sum_{n>0} \chi'^i_n  f_n(z,\bar z),\quad
\chi^{\bar i}=\chi^{\bar i}_0 +\displaystyle \sum_{n>0} \chi'^{\bar i}_n \bar f_n(z,\bar z), 
\label{ex2}
\end{align}
\begin{align}
\psi^i_{\bar z}=\displaystyle \sum_{n>0} \psi'^i_n \frac{1}{\sqrt{E_{n}}}{\d}_{\bar{z}} f_n(z,\bar z),\quad
\psi^{\bar i}_z=\displaystyle \sum_{n>0} \psi'^{\bar i}_n \frac{1}{\sqrt{E_{n}}}\d_{z} \bar f_n(z,\bar z).
\label{ex3}
\end{align}
We set the volume of $\mathbb{C}\mathrm{P^1}$ to 1.
Since $\triangle$ is Hermitian, $\{f_{0}(z,\bar{z})=1, f_1(z,\bar{z}),f_{2}(z,\bar{z}),\cdots\}$ can be considered as orthonormal basis of $\Omega^{(0,0)}$. 
\begin{align}
(f_{n},f_{m}):=\int_{\mathbb{C}\mathrm{P^1}}  \bar f_n(z,\bar z)f_m(z,\bar z) dz\wedge d\bar z= \delta_{n,m}\quad (n,m\geq 0).
\label{ofs1}
\end{align}
Since $(f_{n},(\bar{\d}^{\dag}\bar{\d}+\bar{\d}\bar{\d}^{\dag})f_{m})=E_{m}\delta_{n,m}=(f_{n},\bar{\d}^{\dag}\bar{\d} f_{m})=(\bar{\d}f_{n},\bar{\d}f_{m})$,
we obtain
\begin{align}
\int_{\mathbb{C}\mathrm{P^1}}  \d_{z}\bar f_n(z,\bar z)\d_{\bar{z}}f_m(z,\bar z) dz\wedge d\bar z= E_{m}\delta_{n,m}\quad (n,m>0). 
\label{ofs2}
\end{align}
This forces us to adopt $\{\frac{1}{\sqrt{E_{n}}}\d_{z}f_n(z,\bar{z})\;|\;n>0\}$ and $\frac{1}{\sqrt{E_{n}}}\d_{\bar{z}}f_n(z,\bar{z})\;|\;n>0\}$
as expansion basis of $\psi$. 
Therefore, integration measure for path-integral is given by
\ba
\mathfrak{D}\phi\mathfrak{D}\chi\mathfrak{D}\psi&=&\biggl(\prod_{i=1}^{m}d\phi_{0}^{i}d\phi_{0}^{\bar{i}} d\chi_{0}^{i}d\chi_{0}^{\bar{i}}\biggr)
\biggl(\prod_{i=1}^{m}\prod_{n=1}^{\infty}\frac{d\phi_{n}^{\prime i}d\phi_{n}^{\prime \bar{i}}}{2\pi i}d\chi_{n}^{\prime i}d\chi_{n}^{\prime 
\bar{i}}d\psi_{n}^{\prime i}d\psi_{n}^{\prime \bar{i}}\biggr),\no\\
&=&d\phi_{0}d\chi_{0}\mathfrak{D}\phi'\mathfrak{D}\chi' \mathfrak{D}\psi'.
\ea

\subsection{The case when $E=T^{\prime}M$}
In this subsection, we discuss the case when  the vector bundle $E$ equals $T^{\prime}M$ and the zero set of $K$ is given by a finite set of discrete points $\{p_{1},\cdots,p_{N}\}$. In this case, the action $\Lambda$ on $T^{\prime}M$ is given by $\theta(sK):Y\to [sK,Y]$.
On the zero set of $K$, it is explicitly given as follows.
\begin{align}
\theta(sK)(\frac{\partial}{\partial z^j}) 
= [ sK^i \frac{\partial}{\partial z^i} , \frac{\partial}{\partial z^j}] 
=-s \partial_j K^i \frac{\partial}{\partial z^i}.
\end{align}
Hence we set $\Lambda^i_j = -\partial_j K^i$ in this subsection.

\subsubsection{Explicit Construction of the BRST-closed Observable}

First, we construct the observable $\varphi$. For this purpose, we have only to determine the explicit form of $A^a_b$.
Since the local holomorphic frame $\{e_a\}$ is given by $\{\frac{\partial}{\partial z^a}\}$, we \textcolor{red}{}do not distinguish subscripts of local frame from ones of local coordinate. Then canonical connection becomes $\tilde{\nabla} \frac{\partial}{\partial z^i} =\Gamma^j_{i k} dz^k \frac{\partial}{\partial z^j}$. Then, we obtain
\begin{align*}
&i(sK)\tilde{\nabla} \frac{\partial}{\partial z^i} =sK^li(dz^l)\Gamma^j_{i k} \chi^k \frac{\partial}{\partial z^j}=sK^l\Gamma^j_{i l} \frac{\partial}{\partial z^j},&\\
&F^i_j \frac{\partial}{\partial z^i} =\bar \partial (\Gamma^i_{j k} dz^k \frac{\partial}{\partial z^i} ) = \partial_{\bar l} \Gamma^i_{j k} 
d\overline{z^{l}}\wedge dz^{k} \frac{\partial}{\partial z^i}  =R^i_{j \bar l k}d\overline{z^{l}}\wedge dz^{k}  \frac{\partial}{\partial z^i}=R^i_{j  k \bar l }dz^{k}\wedge d\overline{z^{l}}   \frac{\partial}{\partial z^i}
\end{align*}
and
\begin{align*}
A^i_j \frac{\partial}{\partial z^i}:=L^i_j \frac{\partial}{\partial z^i} + \frac{i}{2\pi }F^i_j \frac{\partial}{\partial z^i}= s\Lambda^i_j \frac{\partial}{\partial z^i}-i(sK)\tilde{\nabla} \frac{\partial}{\partial z^i} -F^i_j \frac{\partial}{\partial z^i} \\= (-s\partial_j K^i -s\Gamma^{ i }_{ j   \mu } K^\mu  + \frac{ i} {2 \pi }  R^i_{ j  k \bar l  } dz^{k}\wedge d\overline{z^{l}} ) \frac{\partial}{\partial z^i}.
\end{align*}
This is the formula we use in this subsection.

\subsubsection{Expansion of the Lagrangian up to the Second Order}

In the $s\to \infty$
limit, path-integral is localized on neighborhood  of $p_{\alpha}$. Therefore, we use expansion of the fields in (\ref{ex1}), (\ref{ex2}) and 
(\ref{ex3}) with $\phi_{0}=p_{\alpha}$.
 Then we expand the Lagrangian up to the second order of the  expansion variables. In the neighborhood of $p_{\alpha}$, $K$ is expanded in the form: $-{K_{\alpha}}^i_j z^j+\cdots$. We can also assume that $ g_{ i \bar j }= \delta_{i \bar j }$ and $\Gamma_{ij}^{k}=\Gamma_{\bar{i}\bar{j}}^{\bar{k}}=0$.
Therefore, in expanding the Lagrangian, 
we can use the following simplification.
\begin{align}
D_z  &\to \partial_z, & D_{\bar z} &\to \partial_{\bar z}, &  \triangledown_j  &\to \partial_j, &   \triangledown_{\bar j } &\to \partial_{\bar j }. & 
\end{align}
We also have $\triangledown_{\bar \mu}\bar K^{\bar j} = -{\bar{K}}^{\bar j}_{\alpha\bar \mu}+\cdots$.
Then expansion of the Lagrangian up to the second order is given as follows.
\begin{align*}
 (L+V)_{\mbox{\tiny 2nd.}}&:=L_{0}^{\alpha}+{L^{\alpha}}',\\ 
L_0^{\alpha} &:=t \bigl[  \beta s^2\delta_{i \bar{j}} {K_{\alpha}}^i_\mu {\bar{K}}^{\bar{j}}_{\alpha\bar \nu} \phi^\mu_0  \phi^{\bar \nu}_0- s\delta_{i \bar{j}} {\bar{K}}^{\bar{j}}_{\alpha\bar \mu} \chi^{\bar \mu}_0 \chi^i_0 \bigr] ,\\
{L^{\alpha}}'&:=\displaystyle \sum_{n>0} \bigl[  t\delta_{i \bar{j} }  \phi'^i_n\phi'^{\bar j}_n E_{n}+  
\sqrt{t}i \delta_{i \bar{j}}( \psi'^{\bar j}_n  \chi'^i_n +\psi'^i_n \chi'^{\bar j}_n)\sqrt{E_{n}}  \bigr]  \\& + \sum_{n>0} \Bigl \{ t\beta s^2\delta_{i \bar{j}} {K_{\alpha}}^i_\mu {\bar{K}}^{\bar{j}}_{\alpha\bar \nu}\phi'^\mu_n\phi'^{\bar \nu}_n - ts\delta_{i \bar{j}}  {\bar{K}}^{\bar{j}}_{\alpha\bar \mu} \chi'^{\bar \mu}_n  \chi'^i_n -s \beta \delta_{i \bar{j}} {K_{\alpha}}^{i}_\mu \psi'^\mu_n  \psi'^{\bar j}_n  \Bigr \},
\label{fex1}
\end{align*}
where $L_{0}$ is zero mode part and $L'$ is oscillation mode part.

\subsubsection{Evaluation of $\lim_{s\rightarrow \infty}<\varphi>_{0}$}
We represent the correlation function in the following form:
\begin{equation}
\lim_{s\rightarrow \infty}<\varphi>_{0} =  \sum_{\alpha} \lim_{s \to \infty} \int \mathfrak{D}\phi\mathfrak{D}\chi\mathfrak{D}\psi  \, \varphi|_{p_{\alpha}} e^{-L_0^{\alpha}-{L^{\alpha}}'}, \label{e1}
\end{equation}
On $p_{\alpha}$,  $A^i_j = - s\partial_j K^i -s\Gamma^{ i }_{ j   \mu } K^\mu  + \frac{i} {2 \pi }R^i_{ j  k \bar l  }dz^k\wedge d\overline{z^{l}}$ becomes $sK^i_{\alpha j}-\frac{ i} {2 \pi }R^i_{ j  k \bar l  } dz^k\wedge d\overline{z^{l}}$.  
Let $K_{\alpha}$ be $m\times m$ matrix defined by $K^{i}_{\alpha j}$.
Then we have $\varphi|_{p_{\alpha}}=s^{m}\varphi(K_{\alpha}-\frac{1}{s}\frac{i}{2\pi}R|_{p_{\alpha}})=:s^{m}\varphi(p_{\alpha},s)$.  
With this set-up, we evaluate the contribution from $p_{\alpha}$.
First, we integrate oscillation modes.  $\varphi|_{p_{\alpha}}$ does not contain $\psi$ oscillation modes and we neglect the third and higher order terms that contain oscillation modes.
The part of oscillation modes integration is given as follows.
\begin{align*}
\lim_{s \to \infty} \int \mathfrak{D}\phi\mathfrak{D}\chi\mathfrak{D}\psi  \, \varphi|_{p_{\alpha}} e^{-{L^{\alpha}}'}
=\lim_{s \to \infty} \varphi|_{p_{\alpha}} \int  \prod_{i=1}^{m}\prod_{n=1}^{\infty}\frac{d\phi_{n}^{\prime i}d\phi_{n}^{\prime \bar{i}}}{2\pi i}d\chi_{n}^{\prime i}d\chi_{n}^{\prime 
\bar{i}}d\psi_{n}^{\prime i}d\psi_{n}^{\prime \bar{i}}  e^{-{L^{\alpha}}'}.
\end{align*}
At this stage, we transform integration variables in the following way $(n>0)$.
\begin{align}
\phi^{\prime \mu}_n&=\frac{1}{s} \phi^\mu_n & 
\phi^{\prime \bar \mu}_n&=\frac{1}{s} \phi^{\bar \mu}_n & 
\chi^{\prime \mu}_n&=\frac{1}{\sqrt{s}} \chi^\mu_n & 
\chi^{\prime \bar \mu}_n&=\frac{1}{\sqrt{s}} \chi^{\bar \mu}_n, & \\
\psi^{\prime \mu}_n&=\frac{1}{\sqrt{s}} \psi^\mu_n & 
\psi^{\prime \bar \mu}_n&=\frac{1}{\sqrt{s}} \psi^{\bar \mu}_n. & 
\end{align}
Then integral measures of oscillation modes are invariant under the transformation and ${L^{\alpha}}'$ is transformed in the following form.
\begin{align}
&\prod_{i=1}^{m}\prod_{n=1}^{\infty}\frac{d\phi_{n}^{\prime i}d\phi_{n}^{\prime \bar{i}}}{2 \pi i}d\chi_{n}^{\prime i}d\chi_{n}^{\prime 
\bar{i}}d\psi_{n}^{\prime i}d\psi_{n}^{\prime \bar{i}}
=\prod_{i=1}^{m}\prod_{n=1}^{\infty}\frac{d\phi_{n}^{i}d\phi_{n}^{ \bar{i}}}{2\pi i }d\chi_{n}^{ i}d\chi_{n}^{
\bar{i}}d\psi_{n}^{ i}d\psi_{n}^{ \bar{i}}, \\ 
&{L^{\alpha}}':=\displaystyle \sum_{n>0} \bigl[  \frac{t}{s^2}\delta_{i \bar{j} }  \phi^i_n\phi^{\bar j}_n E_{n}+  
\frac{\sqrt{t}i}{s} \delta_{i \bar{j}}( \psi^{\bar j}_n  \chi^i_n +\psi^i_n \chi^{\bar j}_n)\sqrt{E_{n}}  \bigr]  \\& + \sum_{n>0} \Bigl \{ t\beta \delta_{i \bar{j}} {K_{\alpha}}^i_\mu {\bar{K}}^{\bar{j}}_{\alpha\bar \nu}\phi^\mu_n\phi^{\bar \nu}_n - t\delta_{i \bar{j}}  {\bar{K}}^{\bar{j}}_{\alpha\bar \mu} \chi^{\bar \mu}_n  \chi^i_n - \beta \delta_{i \bar{j}} {K_{\alpha}}^{i}_\mu \psi^\mu_n  \psi^{\bar j}_n  \Bigr \}.
\end{align}
We neglect $O(s^{-1})$ part since we take the $s \to \infty$ limit. As a result, integration of oscillation modes is given by
\begin{align}
\lim_{s \to \infty} \varphi|_{p_{\alpha}} &\int  \prod_{i=1}^{m}\prod_{n=1}^{\infty}\frac{d\phi_{n}^{i}d\phi_{n}^{ \bar{i}}}{2\pi i}d\chi_{n}^{ i}d\chi_{n}^{
\bar{i}}d\psi_{n}^{ i}d\psi_{n}^{ \bar{i}}  \notag \\
&\times \exp \Bigl[  \sum_{n>0} \Bigl \{ -t\beta \delta_{i \bar{j}} {K_{\alpha}}^i_\mu {\bar{K}}^{\bar{j}}_{\alpha\bar \nu}\phi^\mu_n\phi^{\bar \nu}_n 
+ t\delta_{i \bar{j}}  {\bar{K}}^{\bar{j}}_{\alpha\bar \mu} \chi^{\bar \mu}_n  \chi^i_n 
+ \beta \delta_{i \bar{j}} {K_{\alpha}}^{i}_\mu \psi^\mu_n  \psi^{\bar j}_n  \Bigr \} \Bigr] \\
&=\lim_{s \to \infty} \varphi|_{p_{\alpha}} \prod_{n=1}^{\infty} \frac{1}{(2\pi i)^m} \int  \prod_{i=1}^{m} d\phi_{n}^{i}d\phi_{n}^{ \bar{i}}d\chi_{n}^{ i}d\chi_{n}^{
\bar{i}}d\psi_{n}^{ i}d\psi_{n}^{ \bar{i}}  \notag \\
&\times \exp \Bigl[  -t\beta \delta_{i \bar{j}} {K_{\alpha}}^i_\mu {\bar{K}}^{\bar{j}}_{\alpha\bar \nu}\phi^\mu_n\phi^{\bar \nu}_n 
+ t\delta_{i \bar{j}}  {\bar{K}}^{\bar{j}}_{\alpha\bar \mu} \chi^{\bar \mu}_n  \chi^i_n 
+ \beta \delta_{i \bar{j}} {K_{\alpha}}^{i}_\mu \psi^\mu_n  \psi^{\bar j}_n   \Bigr].
\end{align}
By using the following integral formulas, 
\begin{align}
\int d\phi_0 \mathrm{exp} \bigl[-uM_{i j }\phi^i_0 \phi ^{\bar j }_0 \bigr] &= ( \frac{-2\pi i }{u})^m (\mathrm{det}M)^{-1}, & \\ 
\int d\chi_0 \mathrm{exp} \bigl[  M_{i \bar  j } \chi^i_0 \chi ^{\bar j }_0 \bigr]&=\mathrm{det}M, &
\end{align}
we proceed as follows.
\begin{align}
\lim_{s \to \infty} \varphi|_{p_{\alpha}} &\prod_{n=1}^{\infty} \frac{1}{(2\pi i)^m} \int  \prod_{i=1}^{m} d\phi_{n}^{i}d\phi_{n}^{ \bar{i}}d\chi_{n}^{ i}d\chi_{n}^{
\bar{i}}d\psi_{n}^{ i}d\psi_{n}^{ \bar{i}}  \notag \\
&\times \exp \Bigl[  -t\beta \delta_{i \bar{j}} {K_{\alpha}}^i_\mu {\bar{K}}^{\bar{j}}_{\alpha\bar \nu}\phi^\mu_n\phi^{\bar \nu}_n 
+ t\delta_{i \bar{j}}  {\bar{K}}^{\bar{j}}_{\alpha\bar \mu} \chi^{\bar \mu}_n  \chi^i_n 
+ \beta \delta_{i \bar{j}} {K_{\alpha}}^{i}_\mu \psi^\mu_n  \psi^{\bar j}_n   \Bigr] \\
&=\lim_{s \to \infty} \varphi|_{p_{\alpha}} \prod_{n=1}^{\infty} \Bigl(\frac{-1}{t \beta} \Bigr)^m \frac{(-t \beta)^m \mathrm{det}(\delta_{i \bar{j}}  {\bar{K}}^{\bar{j}}_{\alpha\bar \mu}) \mathrm{det}(\delta_{i \bar{j}} {K_{\alpha}}^{i}_\mu)} 
{\mathrm{det}(\delta_{i \bar{j}} {K_{\alpha}}^i_\mu {\bar{K}}^{\bar{j}}_{\alpha\bar \nu})} \\
&=\lim_{s \to \infty} \varphi|_{p_{\alpha}}  .
\end{align}
Contribution from oscillation modes turn out to be 1. Next, we calculate integral of zero mode part.
\begin{align}
&\lim_{s \to \infty} \int d\phi_0 d\chi_0  \, s^{m}\varphi(p_{\alpha},s) e^{-L_0^{\alpha}} \notag \\
&=\lim_{s \to \infty} s^{m}\varphi(p_{\alpha},s)\frac{1}{\pi^{m}} \int d\phi_0 d\chi_0  \,  \mathrm{exp} \bigl[  -t \bigl( s^2 \beta \delta_{i \bar{j}} K^i _{\alpha\mu }\bar{K}^{\bar{j}} _{\alpha\bar l } \phi^\mu_0 \phi ^{\bar l }_0 - s  \delta_{i \bar{j}} \bar{K}^{\bar{j}}_{\alpha \bar {\mu}} \chi^{\bar {\mu}}_0  \chi^i_0   \bigr)   \bigr]  \notag   \\
&=\lim_{s \to \infty} s^{m}\varphi(p_{\alpha},s) \int d\phi_0 \mathrm{exp} \bigl[  -t  s^2\beta \delta_{i \bar{j}} K^i _{\alpha\mu} \bar{K}^{\bar{j}} _{\alpha\bar l } \phi^\mu_0 \phi ^{\bar l }_0 \bigr] \\
&\times \int d\chi_0 \, \mathrm{exp} \bigl[ t  s  \delta_{i \bar{j}} \bar{K}^{\bar{j}}_{\alpha \bar {\mu}} \chi^{\bar {\mu}}_0  \chi^i_0   \bigr]  \notag \\
&= \lim_{s \to \infty}  \frac{ s^{m} \varphi(p_{\alpha},s)\cdot(2\pi i ts)^m\det(\delta_{i \bar{j}} \bar{K}^{\bar{j}}_{\alpha \bar {\mu}}) }{ (ts^2 \beta)^m \mathrm{det}( \delta_{i \bar{j}} K^i _{\alpha\mu} \bar{K}^{\bar{j}} _{\alpha\bar l } )}
= \lim_{s \to \infty}\frac{ \varphi(p_{\alpha},s)\mathrm{det}(\delta_{i \bar{j}} \bar{K}^{\bar{j}}_{\alpha \bar {\mu}})}{ \mathrm{det}( \delta_{i \bar{j}} K^i _{\alpha\mu} \bar{K}^{\bar{j}} _{\alpha\bar l } )},   
\end{align}
where we used $\beta=2\pi i$.
Since $\lim_{s\to \infty}\varphi(p_{\alpha},s)=\lim_{s\to \infty}\varphi(K_{\alpha}-\frac{1}{s}\frac{i}{2\pi}R|_{p_{\alpha}})=\varphi(K_{\alpha})$, we obtain
\begin{align}
\lim_{s \to \infty} \int \mathfrak{D}\phi_0  \mathfrak{D}\chi_0  \, \varphi(p) e^{-L_0}
&= \frac{ \varphi(K_{\alpha}) \mathrm{det}(\delta_{i \bar{j}} \bar{K}^{\bar{j}}_{\alpha\bar {\mu}})}{ \mathrm{det}( \delta_{i \bar{j}} K^i _{\alpha\mu} \bar{K}^{\bar{j}} _{\alpha\bar l } )}  \notag  \\
&=\frac{\varphi(K_{\alpha})}{ \mathrm{det} (K^i_{\alpha j}) }.  \label{e2}
\end{align}
Since we already know $\theta|_{p_{\alpha}}=K^i_{\alpha j}$, the above result is rewritten by
\begin{equation}
\lim_{s \to \infty}<\varphi > =  \sum_{\alpha=1}^{N} \frac{\varphi(K_{\alpha})}{ \mathrm{det}(\theta|_{p_{\alpha}}) }.
\end{equation}
By combining Proposition 1 and the result in the $s\to 0$ limit, we obtain the Bott residue formula in the case of this subsection.
\begin{equation}
\underline{\varphi}(T^{\prime}M) [M] =  \sum_{\alpha=1}^{N} \frac{\varphi(K_{\alpha})}{ \mathrm{det}(\theta|_{p_{\alpha}}) }.
\end{equation}
Let us assume that $E$ is a general holomorphic vector bundle on $M$ and that zero set of $K$ is given by a discrete point set
$\{p_{1},\cdots,p_{N}\}$. In the same way as the discussion of this subsection, we can derive the Bott residue formula:
\begin{align}
\underline{\varphi}(E) [M] =\sum_{\alpha=1}^{N}  \frac{ \varphi (\Lambda|_{p_{\alpha}} ) }{ \mathrm{ det}(\theta|_{p_{\alpha}}) } 
\end{align}
This result corresponds to the example given in \cite{ref2}.

\subsection{Derivation in General Case} \label{sss1}
In this subsection,  we derive general case of the Bott residue formula, i.e., zero set of $K$ is given by $\{ N_{\alpha}\}$ 
where $N_{\alpha}$ is a connected compact K$\ddot{\mbox a}$hler submanifold of $M$. 
For simplicity, we focus on one connected component $N:=N_{\alpha}$ in the following discussion. 
We set $\mathrm{codim}_{\mathbb{C}} (N) =\nu$. In the $s\to \infty$ limit, the path integral is localized 
to neighborhood of $N$, we apply expansion given in subsection \ref{measure} around the constant map $\phi_{0}\in N$.   
But one subtlety occurs in this case. 
Since $N$ is a K$\ddot{\mbox a}$hler submanifold of $M$, local coordinates around $\phi_{0}\in N$ can be  taken in the following form:
\begin{align*}
( z^1_\perp , \cdots ,z^\nu _\perp ,z^{\nu+1}_{\parallel} , \cdots ,z^m_\parallel ),
\end{align*}
where points in $N$ is described by the condition $z^1_\perp = \cdots  =z^\nu _\perp =0 $. 
Then fields $\phi$, $\chi$ and $\psi$ are also decomposed into $\phi_{\perp}+\phi_{\parallel}$,  $\chi_{\perp}+\chi_{\parallel}$
and $\psi_{\perp}+\psi_{\parallel}$ respectively. From now on, we use alphabets for $\perp$ directions and 
Greek characters for $\parallel$ directions. Then expansion in subsection \ref{measure} is changed as follows. 
\ba
&&\phi^i_\perp=\phi^i_{\perp 0} +\displaystyle \sum_{n >0} \phi'^i_{\perp n} f_n(z,\bar z),\;\; 
\phi^{\bar i}_\perp=\phi^{\bar i}_{\perp 0} +\displaystyle \sum_{n >0} \phi'^{\bar i}_{\perp n} \bar f_n(z,\bar z),\no\\
&&\chi^i_\perp=\chi^i_{\perp 0} +\displaystyle \sum_{n>0} \chi'^i_{\perp n}  f_n(z,\bar z),\;\; 
\chi^{\bar i}_\perp=\chi^{\bar i}_{\perp 0} +\displaystyle \sum_{n>0} \chi'^{\bar i}_{\perp n} \bar f_n(z,\bar z), \no\\
&&\psi^i_{\bar z \perp}=\displaystyle \sum_{n>0}\frac{1}{\sqrt{E_{n}}} \d_{\bar{z}}\psi'^i_{\perp n} f_n(z,\bar z),\;\; 
\psi^{\bar i}_{z \perp}=\displaystyle \sum_{n>0} \psi'^{\bar i}_{\perp n}\frac{1}{\sqrt{E_{n}}} \d_{z}\bar f_n(z,\bar z),\no\\
\label{fex1}
\ea
\ba
&&\phi^\nu_\parallel=\phi^\nu_{\parallel 0} +\displaystyle \sum_{n>0} \phi'^\nu_{\parallel n} f_n(z,\bar z),\;\; 
\phi^{\bar \nu}_\parallel=\phi^{\bar \nu}_{\parallel 0} +\displaystyle \sum_{n>0} \phi'^{\bar \nu}_{\parallel n} \bar f_n(z,\bar z), \no\\
&&\chi^\nu_\parallel=\chi^\nu_{\parallel 0}  +\displaystyle \sum_{n>0} \chi'^\nu_{\parallel n}  f_n(z,\bar z),\;\; 
\chi^{\bar \nu}_\parallel=\chi^{\bar \nu}_{\parallel 0}+\displaystyle \sum_{n>0} \chi'^{\bar \nu}_{\parallel n} \bar f_n(z,\bar z) ,\no\\
&&\psi^\nu_{\bar z \parallel}=\displaystyle \sum_{n>0} \psi'^\nu_{\parallel n}\frac{1}{\sqrt{E_{n}}}  \d_{\bar{z}}f_n(z,\bar z),\;\;   
\psi^{\bar \nu}_{z \parallel}=\displaystyle \sum_{n>0} \psi'^{\bar \nu}_{\parallel n}\frac{1}{\sqrt{E_{n}}}  \d_{z}\bar f_n(z,\bar z).\no\\
\label{fex2}
\ea
In other words, we decompose the integration measure as follows.
\begin{align*}
\int_N \mathfrak{D} \phi_{\parallel 0} \mathfrak{D} \chi_{\parallel 0} \int \mathfrak{D} \phi'_\parallel \mathfrak{D} \chi'_\parallel \mathfrak{D} \psi' _\parallel \int_{N_\perp}  \mathfrak{D} \phi_{\perp 0} \mathfrak{D} \chi_{\perp 0} \int \mathfrak{D} \phi'_\perp \mathfrak{D} \chi'_\perp \mathfrak{D} \psi'_\perp.
\end{align*} 

\subsubsection{Expansion of $L$ up to the Second Order}
By using the orthonormal relation (\ref{ofs1}) and (\ref{ofs2}), expansion $L$ up to the second order is guven by
\begin{align}
L=&
 \displaystyle \sum_{n>0} \Bigl[  
t{E_n}\delta_{i \bar j}  \phi'^i_{\perp n} \phi'^{\bar j}_{\perp n} 
+  \sqrt{t}i\sqrt{E_{n}} \delta_{i \bar j} \psi'^{\bar j}_{\perp n} \chi'^i_{\perp n}
+\sqrt{t}i\sqrt{E_{n}}\delta_{i \bar j}  \psi'^i_{\perp n}  \chi'^{\bar j}_{\perp n}   \notag \\
&+t {E_n}\delta_{\iota \bar \tau}   \phi'^\iota_{\parallel n} \phi'^{\bar \tau}_{\parallel n}   
+  \sqrt{t}i\sqrt{E_{n}} \delta_{\iota \bar \tau}  \psi'^{\bar \tau}_{\parallel n} \chi'^\iota_{\parallel n} 
+\sqrt{t}i\sqrt{E_{n}}\delta_{\iota \bar \tau}  \psi'^\iota_{\parallel n} \chi'^{\bar \tau}_{\parallel n} 
\bigr],
\end{align}
where we used local coordinates that make $g_{i\bar{j}}$ and $\Gamma_{ij}^{k}(\Gamma_{\bar{i}\bar{j}}^{\bar{k}})$ into $\delta_{i\bar{j}}$ and $0$ respectively. 
\subsubsection{Expansion of the Potential $V$}
In this subsection, we expand  the potential term $V$ around neighborhood of $p \in N $. 
\begin{align}
V= \int_{\mathbb{C} \mathrm{P^1}} dz d\bar{z} \bigl[  ts^2\beta g_{I \bar{J}} K^I \bar{K}^{\bar{J}} + tsg_{I \bar{J}} \triangledown_{\bar{M}} \bar{K}^{\bar{J}} \chi^{\bar {M}} \chi^I + s g_{I \bar{J}} \triangledown_{\mu} K^{I}  \psi^{M}_{\bar{z}}  \psi^{\bar{J}}_z \bigr].
\end{align}
where subscripts $I,J,M,\cdots$ run through all the directions $1,2,\cdots,m$.

First, we consider expansion of the second term of $V$.
Let us consider the following Taylor expansion around $p$ .
We use the coordinate system that satisfies 
\begin{align}
&g_{I \bar J} (p)=\delta_{I \bar J}+\cdots\; (g_{i\bar{j}}(p)=\delta_{i\bar{j}}, g_{\mu\bar{\nu}}(p)=\delta_{\mu\bar{\nu}},g_{i\bar{\nu}}(p)=0), \\
&\Gamma^{\bar J}_{\bar I \bar M} =R^{\bar J}_{\bar I L \bar M} \phi^L+\partial_{\bar N} \Gamma^{\bar J}_{\bar I \bar M}(p) \phi^{\bar N}+\cdots.
\end{align} 
Moreover, $p$ is in the zero set of $K$, we can use $K^i =- K^i_l z^l_\perp+\cdots$.  We neglect second order or higher term each.
\begin{align*}
s g_{I \bar J } \triangledown _{\bar M } \bar K^{\bar J }
 &=s g_{I \bar J } (\partial_{\bar M} \bar K^{\bar J }+\Gamma^{\bar J}_{\bar M \bar l} \bar K^{\bar l } )\\
  &=-s\delta_{i \bar j}\bar K^{\bar j }_{\bar m}
 -sR_{I \bar l N \bar M} \bar K^{\bar l }_k \phi^{\bar k}_\perp  \phi^N 
 -s\delta_{I \bar J} \partial_{\bar N} \Gamma^{\bar J}_{\bar I \bar M}  \bar K^{\bar l }_k \phi^{\bar k}_\perp \phi^{\bar N}  \cdots \\
 &=-s\delta_{i \bar j}\bar K^{\bar j }_{\bar m}
 -sR_{I \bar l \mu \bar M} \bar K^{\bar l }_k \phi^{\bar k}_\perp  \phi^\mu_\parallel
 -sR_{I \bar j l \bar M} \bar K^{\bar j }_k \phi^{\bar k}_\perp  \phi^l_\perp \\
& -s\delta_{I \bar J} \partial_{\bar N} \Gamma^{\bar J}_{\bar I \bar M}  \bar K^{\bar l }_k \phi^{\bar k}_\perp \phi^{\bar N} +\cdots.
 \end{align*}
Since the last term \textcolor{red}{}does not contain holomorphic part of $\phi$ and contain $\bar{\phi}_\perp$, it does not contribute to Gaussian integration 
of $\phi_{\perp 0}$ that will be done later.  Hence we neglect this term. Then we obtain
\begin{align}
 sg_{I \bar J } \triangledown _{\bar M } \bar K^{\bar J } 
 = -s\delta_{i \bar j } \bar K^{\bar j}_{\bar m} 
 -sR_{I \bar j \mu  \bar M} \bar K^{\bar j}_{\bar m} \phi^{\bar m }_\perp \phi^\mu_\parallel -sR_{I \bar j l \bar M} \bar K^{\bar j}_{\bar m} \phi^{\bar m }_\perp \phi^l_\perp \label{x1},
 \end{align}
and
\ba
s g_{I \bar J } \triangledown _{\bar M } \bar K^{\bar J } \chi^{\bar M } \chi^I &=& -s\delta_{i \bar j } \bar K^{\bar j}_{\bar m} \chi^{\bar m}_\perp  \chi^i_\perp  -sR_{I \bar j \mu  \bar M} \bar K^{\bar j}_{\bar m} \phi^{\bar m }_\perp \phi^\mu_\parallel \chi^{\bar M } \chi^I 
\no\\
&&-sR_{I \bar j l \bar M} \bar K^{\bar j}_{\bar m} \phi^{\bar m }_\perp \phi^l_\perp  \chi^{\bar M } \chi^I .
\label{ve1}
\ea
Since $\chi=\chi_\parallel +\chi_\perp$, we decompose 
\begin{equation*}
\chi^{\bar M } \chi^I= \chi^{\bar \mu}_\parallel \chi^\iota_\parallel+ \chi^{\bar m}_\perp \chi^\iota_\parallel+ \chi^{\bar \mu}_\parallel \chi^i_\perp + \chi^{\bar m}_\perp \chi^i_\perp .
\end{equation*}
Then the part that corresponds to  (\ref{ve1}) is rewritten as follows.
\begin{align*}
&\int_{\mathbb{C}\mathrm{P^1}} dz d\bar z \Bigl \{s g_{I \bar J } \triangledown _{\bar M } \bar K^{\bar J } \chi^{\bar M } \chi^I \Bigr \} \\
&= \int_{\mathbb{C}\mathrm{P^1}} dz d\bar z \Bigl \{
-s\delta_{i \bar j } \bar K^{\bar j}_{\bar m} \chi^{\bar m}_\perp  \chi^i_\perp  
-sR_{I \bar j \mu  \bar M} \bar K^{\bar j}_{\bar m} \phi^{\bar m }_\perp \phi^\mu_\parallel \chi^{\bar M } \chi^I  \\
&-sR_{I \bar j l \bar M} \bar K^{\bar j}_{\bar m} \phi^{\bar m }_\perp \phi^l_\perp  \chi^{\bar M } \chi^I \Bigr \} .
\end{align*}
Then we use the expansion (\ref{fex1}) and (\ref{fex2}).
\begin{align*}
&\int_{\mathbb{C}\mathrm{P^1}} dz d\bar z \Bigl \{
-s\delta_{i \bar j } \bar K^{\bar j}_{\bar m} \chi^{\bar m}_\perp  \chi^i_\perp \Bigr \}\\
&=-s\delta_{i \bar j } \bar K^{\bar j}_{\bar m} \chi^{\bar m}_{\perp 0} \chi^i_{\perp 0}-s \delta_{i \bar j } \bar K^{\bar j}_{\bar m}\displaystyle \sum_{n \in \mathbb{Z} (n \neq 0)}   \chi'^{\bar m}_{\perp n} \chi'^i_{\perp n},\\
&\int_{\mathbb{C}\mathrm{P^1}} dz d\bar z \Bigl \{
-sR_{I \bar j \mu  \bar M} \bar K^{\bar j}_{\bar m} \phi^{\bar m }_\perp \phi^\mu_\parallel \chi^{\bar M } \chi^I  \Bigr \}\\
&= \int_{\mathbb{C}\mathrm{P^1}} dz d\bar z \Bigl \{
-sR_{I \bar j \mu  \bar M} \bar K^{\bar j}_{\bar m} \{\phi^{\bar m}_{\perp 0} +\displaystyle \sum_{n \in \mathbb{Z} (n \neq 0)} \phi'^{\bar m}_{\perp n} \bar f_n(z,\bar z)\} \phi^\mu_{\parallel 0} \chi^{\bar M } \chi^I  \Bigr \}\\
&= \int_{\mathbb{C}\mathrm{P^1}} dz d\bar z \Bigl \{
-sR_{\iota \bar j \mu  \bar \nu} \bar K^{\bar j}_{\bar m} \phi^{\bar m}_{\perp 0} \phi^\mu_{\parallel 0}  \chi^{\bar \nu}_\parallel \chi^\iota_\parallel
 -sR_{\iota \bar j \mu  \bar l} \bar K^{\bar j}_{\bar m} \phi^{\bar m}_{\perp 0} \phi^\mu_{\parallel 0}\chi^{\bar l}_\perp \chi^\iota_\parallel \\
& -sR_{i \bar j \mu  \bar \nu} \bar K^{\bar j}_{\bar m} \phi^{\bar m}_{\perp 0} \phi^\mu_{\parallel 0}\chi^{\bar \nu}_\parallel \chi^i_\perp 
 -sR_{i \bar j \mu  \bar l} \bar K^{\bar j}_{\bar m} \phi^{\bar m}_{\perp 0} \phi^\mu_{\parallel 0}\chi^{\bar l}_\perp \chi^i_\perp   \Bigr\}\notag \\
&= -s R_{\iota \bar j \mu  \bar \nu} \bar K^{\bar j}_{\bar m} \phi^{\bar m}_{\perp 0} \phi^\mu_{\parallel 0}  \chi^{\bar \nu}_{\parallel 0} \chi^\iota_{\parallel 0}
 -s R_{\iota \bar j \mu  \bar l} \bar K^{\bar j}_{\bar m} \phi^{\bar m}_{\perp 0} \phi^\mu_{\parallel 0}\chi^{\bar l}_{\perp 0} \chi^\iota_{\parallel 0} \\
& -s R_{i \bar j \mu  \bar \nu} \bar K^{\bar j}_{\bar m} \phi^{\bar m}_{\perp 0} \phi^\mu_{\parallel 0}\chi^{\bar \nu}_{\parallel 0} \chi^i_{\perp 0} 
 -sR_{i \bar j \mu  \bar l} \bar K^{\bar j}_{\bar m} \phi^{\bar m}_{\perp 0} \phi^\mu_{\parallel 0}\chi^{\bar l}_{\perp 0} \chi^i_{\perp 0}, \\
&\int_{\mathbb{C}\mathrm{P^1}} dz d\bar z \Bigl \{
-sR_{I \bar j l \bar M} \bar K^{\bar j}_{\bar m} \phi^{\bar m }_\perp \phi^l_\perp  \chi^{\bar M } \chi^I \Bigr \} \\
&=\int_{\mathbb{C}\mathrm{P^1}} dz d\bar z \Bigl \{
-sR_{I \bar j l \bar M} \bar K^{\bar j}_{\bar m} \phi^{\bar m }_{\perp 0} \phi^l_{\perp 0}  \chi^{\bar M } \chi^I \Bigr \} \\
&=-s R_{\iota \bar j l \bar \mu} \bar K^{\bar j}_{\bar m} \phi^{\bar m }_{\perp 0} \phi^l_{\perp 0}  \chi^{\bar \mu}_{\parallel 0} \chi^\iota_{\parallel 0}
- s R_{\iota \bar j l \bar n} \bar K^{\bar j}_{\bar m} \phi^{\bar m }_{\perp 0} \phi^l_{\perp 0}\chi^{\bar n}_{\perp 0} \chi^\iota_{\parallel 0} \\
&- s R_{i \bar j l \bar \mu} \bar K^{\bar j}_{\bar m} \phi^{\bar m }_{\perp 0} \phi^l_{\perp 0}\chi^{\bar \mu}_{\parallel 0} \chi^i_{\perp 0} 
- sR_{i \bar j l \bar n} \bar K^{\bar j}_{\bar m} \phi^{\bar m }_{\perp 0} \phi^l_{\perp 0}\chi^{\bar n}_{\perp 0} \chi^i_{\perp 0}. 
\end{align*}
In this expansion, we neglect third and higher terms that contain oscillation modes. 
Next, we expand the first term by using the same rule.
\begin{align*}
&\int_{\mathbb{C}\mathrm{P^1}} dz d\bar z \Bigl \{s^2 \beta g_{ i \bar j } K^i \bar K^{\bar j} \Bigr \} 
=\int_{\mathbb{C}\mathrm{P^1}} dz d\bar z \Bigl \{s^2 \beta \delta_{ i \bar j } K^i_m \bar K^{\bar j}_{\bar l} \phi^m_\perp \phi^{\bar l}_\perp \Bigr \} \\
&=\int_{\mathbb{C}\mathrm{P^1}} dz d\bar z \Bigl \{s^2 \beta \delta_{ i \bar j } K^i_m \bar K^{\bar j}_{\bar l} \{\phi^m_{\perp 0} +\displaystyle \sum_{n \in \mathbb{Z} (n \neq 0)} \phi'^m_{\perp n} f_n(z,\bar z)\} \\ &\times \{\phi^{\bar l}_{\perp 0} +\displaystyle \sum_{n \in \mathbb{Z} (n \neq 0)} \phi'^{\bar l}_{\perp n} \bar f_n(z,\bar z)\} \Bigr \} \\
&=s^2 \beta \delta_{ i \bar j } K^i_m \bar K^{\bar j}_{\bar l} \phi^m_{\perp 0} \phi^{\bar l}_{\perp 0} +\displaystyle \sum_{n \in \mathbb{Z} (n \neq 0)} s^2 \beta \delta_{ i \bar j } K^i_m \bar K^{\bar j}_{\bar l} \phi'^m_{\perp n} \phi'^{\bar l}_{\perp n} .
\end{align*}
We can neglect the third term by applying tha same rule for expansion. 
As a result, we obtain the following form.
\begin{align*}
&\int_{\mathbb{C}\mathrm{P^1}} dz d\bar z \Bigl \{s^2 \beta g_{ i \bar j } K^i \bar K^{\bar j} +s g_{I \bar J } \triangledown _{\bar M } \bar K^{\bar J } \chi^{\bar M } \chi^I \Bigr \} \\
&=s^2  \beta \delta_{ i \bar j } K^i_m \bar K^{\bar j}_{\bar l} \phi^m_{\perp 0} \phi^{\bar l}_{\perp 0} +
\displaystyle \sum_{n>0} s^2 \beta \delta_{ i \bar j } K^i_m \bar K^{\bar j}_{\bar l} \phi'^m_{\perp n} \phi'^{\bar l}_{\perp n}-s \delta_{i \bar j } \bar K^{\bar j}_{\bar m} \chi^{\bar m}_{\perp 0} \chi^i_{\perp 0} \\
&-s \delta_{i \bar j } \bar K^{\bar j}_{\bar m}\displaystyle \sum_{n>0}   \chi'^{\bar m}_{\perp n} \chi'^i_{\perp n}
-s R_{\iota \bar j \mu  \bar \nu} \bar K^{\bar j}_{\bar m} \phi^{\bar m}_{\perp 0} \phi^\mu_{\parallel 0}  \chi^{\bar \nu}_{\parallel 0} \chi^\iota_{\parallel 0}
-s R_{\iota \bar j \mu  \bar l} \bar K^{\bar j}_{\bar m} \phi^{\bar m}_{\perp 0} \phi^\mu_{\parallel 0}\chi^{\bar l}_{\perp 0} \chi^\iota_{\parallel 0}\\ 
&-s R_{i \bar j \mu  \bar \nu} \bar K^{\bar j}_{\bar m} \phi^{\bar m}_{\perp 0} \phi^\mu_{\parallel 0}\chi^{\bar \nu}_{\parallel 0} \chi^i_{\perp 0} 
-s R_{i \bar j \mu  \bar l} \bar K^{\bar j}_{\bar m} \phi^{\bar m}_{\perp 0} \phi^\mu_{\parallel 0}\chi^{\bar l}_{\perp 0} \chi^i_{\perp 0} \\
 &-s R_{\iota \bar j l \bar \mu} \bar K^{\bar j}_{\bar m} \phi^{\bar m }_{\perp 0} \phi^l_{\perp 0}  \chi^{\bar \mu}_{\parallel 0} \chi^\iota_{\parallel 0}
- s R_{\iota \bar j l \bar n} \bar K^{\bar j}_{\bar m} \phi^{\bar m }_{\perp 0} \phi^l_{\perp 0}\chi^{\bar n}_{\perp 0} \chi^\iota_{\parallel 0} 
\\
&- s R_{i \bar j l \bar \mu} \bar K^{\bar j}_{\bar m} \phi^{\bar m }_{\perp 0} \phi^l_{\perp 0}\chi^{\bar \mu}_{\parallel 0} \chi^i_{\perp 0} 
- s R_{i \bar j l \bar n} \bar K^{\bar j}_{\bar m} \phi^{\bar m }_{\perp 0} \phi^l_{\perp 0}\chi^{\bar n}_{\perp 0} \chi^i_{\perp 0} .
\end{align*}
From the above result, oscillation mode part is the same as the one in the previous subsection.
So, new things that we have to consider is integration of zero mode part.
We summarize the result of expansion of $L+V$ in the following form.
\begin{align}
L+V&=L_0+L'_\parallel +L'_\perp, 
\end{align}
\begin{align}
L_0&:=t\Bigl [s^2  \beta \delta_{ i \bar j } K^i_m \bar K^{\bar j}_{\bar l} \phi^m_{\perp 0} \phi^{\bar l}_{\perp 0}
-s \delta_{i \bar j } \bar K^{\bar j}_{\bar m} \chi^{\bar m}_{\perp 0} \chi^i_{\perp 0} \notag \\
& -s R_{\iota \bar j \mu  \bar \nu} \bar K^{\bar j}_{\bar m} \phi^{\bar m}_{\perp 0} \phi^\mu_{\parallel 0}  \chi^{\bar \nu}_{\parallel 0} \chi^\iota_{\parallel 0} 
-s R_{\iota \bar j \mu  \bar l} \bar K^{\bar j}_{\bar m} \phi^{\bar m}_{\perp 0} \phi^\mu_{\parallel 0}\chi^{\bar l}_{\perp 0} \chi^\iota_{\parallel 0} \notag \\
&-s R_{i \bar j \mu  \bar \nu} \bar K^{\bar j}_{\bar m} \phi^{\bar m}_{\perp 0} \phi^\mu_{\parallel 0}\chi^{\bar \nu}_{\parallel 0} \chi^i_{\perp 0} 
-s R_{i \bar j \mu  \bar l} \bar K^{\bar j}_{\bar m} \phi^{\bar m}_{\perp 0} \phi^\mu_{\parallel 0}\chi^{\bar l}_{\perp 0} \chi^i_{\perp 0} \notag \\
& -s R_{\iota \bar j l \bar \mu} \bar K^{\bar j}_{\bar m} \phi^{\bar m }_{\perp 0} \phi^l_{\perp 0}  \chi^{\bar \mu}_{\parallel 0} \chi^\iota_{\parallel 0}
- s R_{\iota \bar j l \bar n} \bar K^{\bar j}_{\bar m} \phi^{\bar m }_{\perp 0} \phi^l_{\perp 0}\chi^{\bar n}_{\perp 0} \chi^\iota_{\parallel 0} \notag \\
&- s R_{i \bar j l \bar \mu} \bar K^{\bar j}_{\bar m} \phi^{\bar m }_{\perp 0} \phi^l_{\perp 0}\chi^{\bar \mu}_{\parallel 0} \chi^i_{\perp 0} 
- s R_{i \bar j l \bar n} \bar K^{\bar j}_{\bar m} \phi^{\bar m }_{\perp 0} \phi^l_{\perp 0}\chi^{\bar n}_{\perp 0} \chi^i_{\perp 0}  \Bigr ],
\label{lex0}
\end{align}
\begin{align}
L'_\parallel&:= \sum_{n>0} \Bigl [
tE_n\delta_{\iota \bar \tau}   \phi'^\iota_{\parallel n} \phi'^{\bar \tau}_{\parallel n}   
+  \sqrt{tE_{n}}i  \delta_{\iota \bar \tau}  \psi'^{\bar \tau}_{\parallel n} \chi'^\iota_{\parallel n} 
+ \sqrt{tE_{n}}i \delta_{\iota \bar \tau}  \psi'^\iota_{\parallel n} \chi'^{\bar \tau}_{\parallel n} 
\Bigr],
\label{lex1}
\end{align}
\begin{align}
L'_\perp &:= \sum_{n>0} \Bigl \{ tE_{n}\delta_{i \bar j}  \phi'^i_{\perp n} \phi'^{\bar j}_{\perp n} 
+  \sqrt{tE_{n}}i \delta_{i \bar j} \psi'^{\bar j}_{\perp n} \chi'^i_{\perp n}
+ \sqrt{tE_{n}}i \delta_{i \bar j}  \psi'^i_{\perp n}  \chi'^{\bar j}_{\perp n} \notag \\
&+t\beta s^2\delta_{i \bar{j}} K^i_\mu \bar{K}^{\bar{j}}_{\bar \nu}\phi'^\mu_{\perp n} \phi'^{\bar \nu}_{\perp n} - s\delta_{i \bar{j}}  \bar{K}^{\bar{j}}_{\bar \mu} \chi'^{\bar \mu}_{\perp n}  \chi'^i_{\perp n} - s\beta \delta_{i \bar{j}} K^{i}_\mu \psi'^\mu_{\perp n}  \psi'^{\bar j}_{\perp n}  \Bigr \}.
\label{lex2}
\end{align}

\subsubsection{Evaluation of $\lim_{s\rightarrow \infty}<\varphi>_{0}$}
First, we  decompose $L_0$ into $L_{\perp 0}+L_{\parallel 0}$.
For this purpose, we transform variables in the following way ($n>0$).
\begin{align}
\phi^\iota _{\parallel 0}& =\phi''^{\iota} _{\parallel 0},&
\phi^{\bar \iota} _{\parallel 0} &= \phi''^{\bar \iota} _{\parallel 0},&
\phi^i _{\perp 0} &=\frac{1}{s} \phi''^i _{\perp 0}, &
\phi^{\bar i} _{\perp 0}& =\frac{1}{s} \phi''^{\bar i} _{\perp 0},& 
\notag \\
\chi^\iota_{\parallel 0} &= \sqrt{s}  \chi''^\iota_0, &
\chi^{\bar \iota}_{\parallel 0} &= \sqrt{s}  \chi''^{\bar \iota}_0, & 
\chi^i_{\perp 0} &= \frac{1}{\sqrt{s}} \chi''^i_{\perp 0}, &
\chi^{\bar i}_{\perp 0} &= \frac{1}{\sqrt{s}} \chi''^{\bar i}_{\perp 0}, & \\
\phi'^\iota _{\parallel n}& =\phi'''^{\iota} _{\parallel n},&
\phi'^{\bar \iota} _{\parallel n} &= \phi'''^{\bar \iota} _{\parallel n},&
\phi'^i _{\perp n} &=\frac{1}{s} \phi'''^i _{\perp n}, &
\phi'^{\bar i} _{\perp n}& =\frac{1}{s} \phi''^{\bar i} _{\perp 0},& 
\notag \\
\chi'^\iota_{\parallel n} &=  \chi'''^\iota_{\parallel n}, &
\chi'^{\bar \iota}_{\parallel n} &= \chi'''^{\bar \iota}_{\parallel n}, & 
\chi'^i_{\perp n} &= \frac{1}{\sqrt{s}} \chi''^i_{\perp 0}, &
\chi'^{\bar i}_{\perp n} &= \frac{1}{\sqrt{s}} \chi''^{\bar i}_{\perp n}, 
\notag \\
\psi'^\iota _{\parallel n}& =\psi'''^{\iota} _{\parallel n},&
\psi'^{\bar \iota} _{\parallel n} &= \psi'''^{\bar \iota} _{\parallel n},&
\psi'^i _{\perp n} &= \frac{1}{\sqrt{s}}  \psi'''^i _{\perp n}, &
\psi'^{\bar i} _{\perp n}& =\frac{1}{\sqrt{s}}  \psi''^{\bar i} _{\perp 0}.& 
\end{align}
Let us focus on the measure: 
 \begin{align*}
d\phi_{\parallel 0} d\chi_{\parallel 0} =d\phi^{\nu+1}_{\parallel 0} d\phi^{\overline{\nu+1}}_{\parallel 0} \cdots d\phi^{m}_{\parallel 0} d
\phi^{ \overline{m}} _{\parallel 0} d\chi^{\nu+1}_{\parallel 0} d\chi^{\overline{\nu+1}}_{\parallel 0} \cdots  d\chi^{m}_{\parallel 0} 
d\chi^{\overline{m}}_{\parallel 0}. 
 \end{align*}
 Transformation rule of the measure is given by Berezinian as follows.
\begin{align*}
d \phi_{\parallel 0} d\chi_{\parallel 0} =\frac{1}{s^{m-\nu}}d \phi''_{\parallel 0} d \chi''_{\parallel 0}.
\end{align*}
As for the measure:
\begin{align*}
d\phi_{\perp 0} d\chi_{\perp 0} =d\phi^{1}_{\perp 0} d\phi^{\overline{1}}_{\perp 0} \cdots d\phi^\nu _{\perp 0} d\phi^{\overline{ \nu}}_{\perp 0} d\chi^{1}_{\perp 0} d\chi^{\overline{1}}_{\perp 0} \cdots  d\chi^{\nu}_{\perp 0} d\chi^{\overline{ \nu }}_{\perp 0},
\end{align*}
transformation rule is given by
\begin{align*}
d\phi_{\perp 0} d \chi_{\perp 0} =\frac{1}{s^\nu} d\phi''_{\perp 0}d\chi''_{\perp 0}.
\end{align*}
Integral measures of oscillation modes are defined as follows.
\begin{align*}
 \mathfrak{D} \phi'_\parallel \mathfrak{D} \chi'_\parallel \mathfrak{D}\psi'_\parallel=\displaystyle \prod_{n>0} &\frac{1}{(2 \pi i)^{m-\nu}} d\phi'^{\nu+1}_{\parallel n } d\phi'^{\overline{\nu+1}}_{\parallel n} \cdots d\phi'^{m}_{\parallel n} d
\phi'^{ \overline{m}} _{\parallel n} 
d\chi'^{\nu+1}_{\parallel n} d\chi'^{\overline{\nu+1}}_{\parallel n} \cdots  d\chi'^{m}_{\parallel n} 
d\chi'^{\overline{m}}_{\parallel n } \\
&\times d\psi'^{\nu+1}_{\parallel n } d\psi'^{\overline{\nu+1}}_{\parallel n} \cdots d\psi'^{m}_{\parallel n} d
\psi'^{ \overline{m}} _{\parallel n} ,
\end{align*}
\begin{align*}
\mathfrak{D} \phi'_{\perp } \mathfrak{D} \chi'_{\perp } \mathfrak{D} \psi'_{\perp }=\displaystyle \prod_{n>0}\frac{1}{(2 \pi i)^\nu} &d\phi'^{1}_{\perp n} d\phi'^{\overline{1}}_{\perp n } \cdots d\phi'^\nu _{\perp n} d\phi'^{\overline{ \nu}}_{\perp n} 
d\chi'^{1}_{\perp n } d\chi'^{\overline{1}}_{\perp n} \cdots  d\chi'^{\nu}_{\perp n } d\chi'^{\overline{ \nu }}_{\perp n} \\
&\times d\psi'^{1}_{\perp n} d\psi'^{\overline{1}}_{\perp n } \cdots d\psi'^\nu _{\perp n} d\psi'^{\overline{ \nu}}_{\perp n}. 
\end{align*}
These are invariant under the transformation.
\begin{align*}
\mathfrak{D}\phi'_\parallel \mathfrak{D} \chi'_\parallel \mathfrak{D}\psi'_\parallel=\mathfrak{D} \phi'''_\parallel \mathfrak{D} \chi'''_\parallel \mathfrak{D}\psi'''_\parallel,
\end{align*}
\begin{align*}
\mathfrak{D}\phi'_{\perp } \mathfrak{D}\chi'_{\perp } \mathfrak{D}\psi'_{\perp }=\mathfrak{D}\phi'''_{\perp } \mathfrak{D}\chi'''_{\perp } \mathfrak{D}\psi'''_{\perp }.
\end{align*}
In sum, transformation of the whole integral measure is given by
\begin{align*}
&\int_N \mathfrak{D} \phi_{\parallel 0} \mathfrak{D} \chi_{\parallel 0} \int \mathfrak{D} \phi'_\parallel \mathfrak{D} \chi'_\parallel \mathfrak{D} \psi' _\parallel \int_{N_\perp}  \mathfrak{D} \phi_{\perp 0} \mathfrak{D} \chi_{\perp 0} \int \mathfrak{D} \phi'_\perp \mathfrak{D} \chi'_\perp \mathfrak{D} \psi'_\perp  \\
&=\frac{1}{ s^m} \int_{N} d\phi''_{\parallel 0} d\chi''_{\parallel 0} \int_{N_{\perp }}  
d\phi''_{\perp 0} d\chi''_{\perp 0} \int \mathfrak{D}\phi'''_\parallel \mathfrak{D}\chi'''_\parallel \mathfrak{D}\psi'''_\parallel \int \mathfrak{D}\phi'''_{\perp }\mathfrak{D}\chi'''_{\perp } \mathfrak{D}\psi'''_{\perp }.
\end{align*}
Let us consider $\varphi$ at $p \in N$. 
By using the above variables, $A^b_a$ is expanded in the following form:
\begin{align*}
 A^b_a=s\Lambda^b_a +\frac{i}{2\pi} {F|_p}^b_{a I \bar J} \chi^I \chi^J
 =s(\Lambda^b_a +\frac{i}{2\pi} F^b_{a \nu \bar \mu} \chi''^\nu_0 \chi''^{\bar \mu}_0 )+\sqrt{s}\{\cdots\} +\cdots+s^{-1}\{ \cdots\}
\end{align*}
Since we neglect the third and higher terms that contain oscillation modes, $F|_p$ does not depend on $\phi_\perp$.
Then we expand $\varphi (p)$ in the form:
 $\varphi (p)=\displaystyle \sum^{2m}_{k=-2m} s^{\frac{k}{2}}\varphi_k(p)$.
Note that $\varphi_{2m}(p)$ is written as $\varphi(\Lambda^b_a +\frac{i}{2\pi} F^b_{a \nu \bar \mu} \chi''^\nu_0 \chi''^{\bar \mu}_0)$. 
We then look back at the expansion of the potential term. Since we neglect terms that have negative powers in $s$, it is represented as follows.
\begin{align*}
L_0
&=t \Bigl \{ \beta \delta_{ i \bar j } K^i_m \bar K^{\bar j}_{\bar l} \phi''^m_{\perp 0} \phi''^{\bar l}_{\perp 0}
-\delta_{i \bar j } \bar K^{\bar j}_{\bar m} \chi''^{\bar m}_{\perp 0} \chi''^i_{\perp 0} \notag \\
& -s R_{\iota \bar j \mu  \bar \nu} \bar K^{\bar j}_{\bar m} \phi''^{\bar m}_{\perp 0} \phi''^\mu_{\parallel 0}  \chi''^{\bar \nu}_{\parallel 0} \chi''^\iota_{\parallel 0} 
-R_{\iota \bar j \mu  \bar l} \bar K^{\bar j}_{\bar m} \phi''^{\bar m}_{\perp 0} \phi''^\mu_{\parallel 0}\chi''^{\bar l}_{\perp 0} \chi''^\iota_{\parallel 0} \notag \\
&-R_{i \bar j \mu  \bar \nu} \bar K^{\bar j}_{\bar m} \phi''^{\bar m}_{\perp 0} \phi''^\mu_{\parallel 0}\chi''^{\bar \nu}_{\parallel 0} \chi^i_{\perp 0} 
 -R_{\iota \bar j l \bar \mu} \bar K^{\bar j}_{\bar m} \phi''^{\bar m }_{\perp 0} \phi''^l_{\perp 0}  \chi''^{\bar \mu}_{\parallel 0} \chi''^\iota_{\parallel 0}
\Bigr\},
\end{align*}
\begin{align}
L'_\parallel&= \sum_{n>0} \Bigl [
tE_n\delta_{\iota \bar \tau}   \phi'''^\iota_{\parallel n} \phi'''^{\bar \tau}_{\parallel n}   
+  \sqrt{tE_{n}}i\delta_{\iota \bar \tau}  \psi'''^{\bar \tau}_{\parallel n} \chi'''^\iota_{\parallel n} 
+ \sqrt{tE_{n}}i\delta_{\iota \bar \tau}  \psi'''^\iota_{\parallel n} \chi'''^{\bar \tau}_{\parallel n} 
\Bigr],\\
L'_\perp &= \sum_{n>0} \Bigl[ t\beta \delta_{i \bar{j}} K^i_m \bar{K}^{\bar{j}}_{l}\phi'''^m_{\perp n} \phi'''^{\bar l}_{\perp n} -t \delta_{i \bar{j}}  \bar{K}^{\bar{j}}_{\bar m} \chi'''^{\bar m}_{\perp n}  \chi'^i_{\perp n} - \beta \delta_{i \bar{j}} K^{i}_m \psi'''^m_{\perp n}  \psi'''^{\bar j}_{\perp n}  \Bigr ].
\end{align}
Let us evaluate the contribution from the component $N$ to $\lim_{s\rightarrow \infty}<\varphi>_{0}$.
First, we integrate oscillation modes in $\parallel$-part. Since each $\varphi_k(p)$ \textcolor{red}{}does not contain  $\psi'''_\parallel$ and $\phi'''_\parallel$,
we have only to perform simple Gaussian integral. 
\begin{align*}
&\int \mathfrak{D}\phi'''_\parallel \mathfrak{D} \chi'''_\parallel \mathfrak{D}\psi'''_\parallel \varphi(p) e^{ -L'_\parallel} \\
&= \varphi'(p) \int \mathfrak{D}\phi'''_\parallel \mathfrak{D} \chi'''_\parallel \mathfrak{D}\psi'''_\parallel \exp \Bigl \{ -
\sum_{n>0} \Bigl [
tE_n\delta_{\iota \bar \tau}   \phi'''^\iota_{\parallel n} \phi'''^{\bar \tau}_{\parallel n} +   \sqrt{tE_{n}}i \delta_{\iota \bar \tau}  \psi'''^{\bar \tau}_{\parallel n} \chi'''^\iota_{\parallel n} 
+  \sqrt{tE_{n}}i \delta_{\iota \bar \tau}  \psi'''^\iota_{\parallel n} \chi'''^{\bar \tau}_{\parallel n} 
\Bigr]
 \Bigr \} \\
&=\varphi'(p) \displaystyle \prod_{n>0}\frac{1}{(2\pi i)^{m-\nu}} \int d\phi'''_{\parallel n} d \chi'''_{\parallel n} d\psi'''_{\parallel n} \\
&\times  \exp \Bigl \{- \Bigl [
tE_n\delta_{\iota \bar \tau}   \phi'''^\iota_{\parallel n} \phi'''^{\bar \tau}_{\parallel n}+ \sqrt{tE_{n}}i \delta_{\iota \bar \tau}  \psi'''^{\bar \tau}_{\parallel n} \chi'''^\iota_{\parallel n} 
+  \sqrt{tE_{n}}i \delta_{\iota \bar \tau}  \psi'''^\iota_{\parallel n} \chi'''^{\bar \tau}_{\parallel n} 
\Bigr]\Bigr \} \\
&=\varphi'(p) \displaystyle \prod_{n>0} \Bigl(\frac{tE_{n}}{tE_n} \Bigr)^{m-\nu}=\varphi'(p) .
\end{align*}
We mean by $\varphi'(p)$ the operator obtained from removing $\chi'''_\parallel$ from $\varphi(p)$.
Next, we integrate oscillation modes in $\perp$-part. We expand $\varphi'(p)=\displaystyle \sum^{2m}_{k=-2m} s^{\frac{k}{2}}\varphi'_k(p),\,\,(\varphi_{m}(p)=\varphi'_{2m}(p)=\varphi(\Lambda^b_a +\frac{i}{2\pi} F^b_{a \nu \bar \mu} \chi''^\nu_0 \chi''^{\bar \mu}_0))$. Then we obtain
\begin{align*}
&\int \mathfrak{D}\phi'''_{\perp } \mathfrak{D}\chi'''_{\perp } \mathfrak{D} \psi'''_{\perp } \varphi'(p) \exp(-L'_\perp)\\
&=\int \mathfrak{D} \phi'''_{\perp } \mathfrak{D} \chi'''_{\perp } \mathfrak{D} \psi'''_{\perp } \varphi'(p) \exp\Bigl\{-\sum_{n>0} \Bigl\{ t\beta \delta_{i \bar{j}} K^i_m \bar{K}^{\bar{j}}_{l}\phi'''^m_{\perp n} \phi'''^{\bar l}_{\perp n} -t \delta_{i \bar{j}}  \bar{K}^{\bar{j}}_{\bar m} \chi'''^{\bar m}_{\perp n}  \chi'^i_{\perp n} \\
&- \beta \delta_{i \bar{j}} K^{i}_m \psi'''^m_{\perp n}  \psi'''^{\bar j}_{\perp n}  \Bigr \} \Bigr \}\\
&=s^{m} \varphi'_{2m}(p) \displaystyle \prod_{n>0}\frac{1}{(2\pi i)^{\nu}} \int d\phi'''_{\perp n} \exp\Bigl\{  
-t\beta \delta_{i \bar{j}} K^i_m \bar{K}^{\bar{j}}_{l}\phi'''^m_{\perp n} \phi'''^{\bar l}_{\perp n} \Bigr \}\\
&\times \int d\chi'''_{\perp n} d\psi'''_{\perp n}  \exp\Bigl\{ t \delta_{i \bar{j}}  \bar{K}^{\bar{j}}_{\bar m} \chi'''^{\bar m}_{\perp n}  \chi'^i_{\perp n} + \beta \delta_{i \bar{j}} K^{i}_m \psi'''^m_{\perp n}  \psi'''^{\bar j}_{\perp n} \Bigr \}\\
&=s^{m} \varphi'_{2m}(p)\prod_{n>0}\Bigl( \frac{  \mathrm{det}\bigl(\delta_{i \bar{j}}  \bar{K}^{\bar{j}}_{\bar m}\bigr) \mathrm{det}\bigl(\delta_{i \bar{j}} K^{i}_l\bigr)}{\mathrm{det} \bigl(\delta_{i \bar{j}} K^i_m \bar{K}^{\bar{j}}_{\bar l} \bigr)}
 \Bigr)+(\mbox{terms of lower power in $s$})\\
&=s^{m} \varphi'_{2m}(p) +(\mbox{terms of lower power in $s$}).
\end{align*}
At this stage, we can represent the contribution from $N$ in the following form.
\begin{align*}
 \int_{N_\alpha} d\phi''_{\parallel 0} d\chi''_{\parallel 0} \varphi'_{2m}(p) \int_{N_{\perp \alpha}}  
d\phi''_{\perp 0} d\chi''_{\perp 0} ~\exp\{ -L_0\}+(\mbox{terms of lower power in $s$}).
\end{align*}
Finally, we integrate out zero modes in $\perp $-part.
\begin{align*}
&\int_{N_{\perp}} d\phi''_{\perp 0} d\chi''_{\perp 0} ~\exp\{ -L_0\}\\
&=\int_{N_{\perp}} d\phi''_{\perp 0}\exp \Bigl[ -t \phi''^m_{\perp 0} \Bigl \{ \beta \delta_{ i \bar j } K^i_m \bar K^{\bar j}_{\bar l} 
 -R_{\iota \bar j m \bar \mu} \bar K^{\bar j}_{\bar l}  \chi''^{\bar \mu}_{\parallel 0} \chi''^\iota_{\parallel 0} \Bigl \} \phi''^{\bar l}_{\perp 0} 
 +t s R_{\iota \bar j \mu  \bar \nu} \bar K^{\bar j}_{\bar m} \phi''^{\bar m}_{\perp 0} \phi''^\mu_{\parallel 0}  \chi''^{\bar \nu}_{\parallel 0} \chi''^\iota_{\parallel 0}  \Bigr] 
\end{align*}
\begin{align*}
&\times \int d\chi''_{\perp 0} \exp \Bigl[ t \Bigl \{ \delta_{i \bar j } \bar K^{\bar j}_{\bar m} \chi''^{\bar m}_{\perp 0} \chi''^i_{\perp 0} 
+R_{\iota \bar j \mu  \bar l} \bar K^{\bar j}_{\bar m} \phi''^{\bar m}_{\perp 0} \phi''^\mu_{\parallel 0}\chi''^{\bar l}_{\perp 0} \chi''^\iota_{\parallel 0} 
+R_{i \bar j \mu  \bar \nu} \bar K^{\bar j}_{\bar m} \phi''^{\bar m}_{\perp 0} \phi''^\mu_{\parallel 0}\chi''^{\bar \nu}_{\parallel 0} \chi^i_{\perp 0} \Bigr\} \Bigr] \\
&=\int_{N_{\perp}} d\phi''_{\perp 0}\exp \Bigl[ -t \phi''^m_{\perp 0} \Bigl \{ \beta \delta_{ i \bar j } K^i_m \bar K^{\bar j}_{\bar l} 
 -R_{\iota \bar j m \bar \mu} \bar K^{\bar j}_{\bar l}  \chi''^{\bar \mu}_{\parallel 0} \chi''^\iota_{\parallel 0} \Bigl \} \phi''^{\bar l}_{\perp 0} 
 +t s R_{\iota \bar j \mu  \bar \nu} \bar K^{\bar j}_{\bar m} \phi''^{\bar m}_{\perp 0} \phi''^\mu_{\parallel 0}  \chi''^{\bar \nu}_{\parallel 0} \chi''^\iota_{\parallel 0}  \Bigr] \\
&\times \Bigl[(-t)^\nu \mathrm{det}\bigl(\delta_{i \bar j } \bar K^{\bar j}_{\bar m}\bigr) +\cdots \\
&+\int d\chi''_{\perp 0} \exp\bigl( 
tR_{\iota \bar j \mu  \bar l} \bar K^{\bar j}_{\bar m} \phi''^{\bar m}_{\perp 0} \phi''^\mu_{\parallel 0}\chi''^{\bar l}_{\perp 0} \chi''^\iota_{\parallel 0} 
+tR_{i \bar j \mu  \bar \nu} \bar K^{\bar j}_{\bar m} \phi''^{\bar m}_{\perp 0} \phi''^\mu_{\parallel 0}\chi''^{\bar \nu}_{\parallel 0} \chi^i_{\perp 0}\bigr) \Bigr].
\end{align*}
Then we preform Gaussian integral of $\exp( -t \phi''^m_{\perp 0} \Bigl \{ \beta \delta_{ i \bar j } K^i_m \bar K^{\bar j}_{\bar l} 
 -R_{\iota \bar j m \bar \mu} \bar K^{\bar j}_{\bar l}  \chi''^{\bar \mu}_{\parallel 0} \chi''^\iota_{\parallel 0} \Bigl \} \phi''^{\bar l}_{\perp 0} )$. Note that terms except for $(t)^\nu \mathrm{det}\bigl(\delta_{i \bar j } \bar K^{\bar j}_{\bar m}\bigr)$ include grassmann variables. Hence by expanding exponential, we only have to consider polynomial correlation function of $\phi_{\perp 0}^{*}$ for these terms.
But the matrix $\Bigl \{ \beta \delta_{ i \bar j } K^i_m \bar K^{\bar j}_{\bar l} 
 -R_{\iota \bar j m \bar \mu} \bar K^{\bar j}_{\bar l}  \chi''^{\bar \mu}_{\parallel 0} \chi''^\iota_{\parallel 0} \Bigl \}$
takes the form $A_{i\bar{j}}$, and these correlation function all vanishes because they only have anti-holomorphic variable $\phi_{\perp 0}^{\bar{m}}$. As a result, we obtain 
\begin{align*}
&\int_{N_{\perp }} d\phi''_{\perp 0} d\chi''_{\perp 0} ~\exp\{ -L_0\}
=\Bigl(\frac{-2\pi i }{t} \Bigr)^\nu \frac{(-t)^\nu \mathrm{det}\{\delta_{i \bar j } \bar K^{\bar j}_{\bar m}\}}{\mathrm{det} \{ \beta \delta_{ i \bar j } K^i_m \bar K^{\bar j}_{\bar l} 
 -R_{\iota \bar j m \bar \mu} \bar K^{\bar j}_{\bar l}  \chi''^{\bar \mu}_{\parallel 0} \chi''^\iota_{\parallel 0}  \} }
\\
&= \frac{(2 \pi i)^\nu \mathrm{det}\{\delta_{i \bar j } \bar K^{\bar j}_{\bar m}\}}{(\beta)^\nu \mathrm{det} \{ \delta_{ i \bar j } K^i_m \bar K^{\bar j}_{\bar l} 
 +\frac{i}{2\pi} R_{\iota \bar j m \bar \mu} \bar K^{\bar j}_{\bar l}  \chi''^{\bar \mu}_{\parallel 0} \chi''^\iota_{\parallel 0}  \} }.
\end{align*}
 From $R_{\iota \bar j m \bar \mu} =R_{m \bar j \iota \bar \mu} =-\delta_{i \bar j} R^i_{m \iota \bar \mu}$,
\begin{align*}
&\int_{N_{\perp}} d\phi''_{\perp 0} d\chi''_{\perp 0} ~\exp\{ -L_0\}
= \frac{1 }{  \mathrm{det} \{ K^i_m +\frac{i}{2\pi} R^i_{m \iota \bar \mu}  \chi''^\iota_{\parallel 0} \chi''^{\bar \mu}_{\parallel 0} \} }.
\end{align*}
We remark $\varphi'_{2m}(p)=\varphi(\Lambda^b_a +\frac{i}{2\pi} F^b_{a \nu \bar \mu} \chi''^\nu_0 \chi''^{\bar \mu}_0)=\underline{\varphi}(\Lambda_\alpha)$. By adding up contributions from all the connected components, 
we obtain  the correlation function in the following form.
\begin{align*}
\displaystyle \lim_{s\rightarrow \infty}<\varphi>_{0}&=
\displaystyle \lim_{s\rightarrow \infty} \displaystyle \sum_{\alpha} \Bigl[\int_{N_\alpha} d\phi''_{\parallel 0} d\chi''_{\parallel 0} \frac{\underline{\varphi}(\Lambda_\alpha) }{ \mathrm{det} \{ K^i_m 
 +\frac{i}{2\pi} R^i_{m \iota \bar \mu}  \chi''^\iota_{\parallel 0} \chi''^{\bar \mu}_{\parallel 0} \} } \\
&+(\mbox{terms of negative power in $s$})\Bigr]\\
&=\displaystyle \sum_{\alpha} \int_{N_\alpha} d\phi''_{\parallel 0} d\chi''_{\parallel 0}  \frac{\underline{\varphi}(\Lambda_\alpha) }{ \mathrm{det} \{ K^i_m 
 +\frac{i}{2\pi} R^i_{m \iota \bar \mu}  \chi''^\iota_{\parallel 0} \chi''^{\bar \mu}_{\parallel 0} \} } .
\end{align*}
Lastly, we rewrite the determinant in denominator. 
$K^i_j$ corresponds to the map $\theta^\nu |_{N} :T'M|_{N}/T'N \to T'M|_{N}/T'N$. 
and $R^{ i }_{  l   \iota \bar \mu  }\chi'^\iota _0 \chi'^{\bar \mu }_0$ is nothing but the curvature $(1,1)$-form of  $T'M|_{N}/T'N$ ( $R^i_l=R^\nu _\alpha$). As a result, we can rewrite the above result into the form:
\begin{align*}
\displaystyle \lim_{s\rightarrow \infty}<\varphi>_{0}
=\displaystyle \sum_{\alpha} \int_{N_\alpha}\frac{\underline{\varphi}(\Lambda_\alpha) }{ \mathrm{det} \{ \theta^\nu_\alpha+\frac{i}{2\pi} R^\nu_\alpha \} }.
\end{align*}
By combining the above result with Proposition 1, we finally obtain the Bott residue formula:
\begin{align*}
\underline{\varphi}(E) [M]
=\displaystyle \sum_{\alpha} \int_{N_\alpha}\frac{\underline{\varphi}(\Lambda_\alpha) }{ \mathrm{det} \{ \theta^\nu_\alpha+\frac{i}{2\pi} R^\nu_\alpha \} }. 
\end{align*}

\appendix
\def\thesection{Appendix\Alph{section}}

\section{Proof of Proposition 1}

We prove the correlation function is independent of parameter $s$. The basic idea comes from \cite{ref3} and \cite{ref7}. Sigma model has two charge for fermion $F_A$ and $F_V$. These charge acts on the operator ${\cal O}_{\omega}$ obtained from 
$(p,q)$-form $\omega$ as follows.
\begin{align}
F_A {\cal O}_{\omega} &= (p+q) {\cal O}_{\omega}, & F_V{\cal O}_{\omega} = (-p+q){\cal O}_{\omega}.&
\end{align}
The symmetry $F_A$ is broken by the potential term. So, observables are graded by $F_V$. However, since $F_A$ is counting total degree  of differential forms, we can use it for taking conjugation of operators (this idea was used in the discussion on Landau-Ginzburg model in \cite{ref7}). Let us consider $e^{\lambda F_A} ~(\lambda \in \mathbb{R})$. 
Then $e^{- \lambda F_A} Q_s e^{\lambda F_A}$ is evaluated as follows.
\begin{align}
e^{- \lambda F_A} Q_s e^{\lambda F_A}{\cal O}_{\omega} =e^{-\lambda} Q_{se^{2\lambda}}{\cal O}_{\omega}. 
\end{align}
Next, we focus on the observable $\varphi_s$. We decompose observable $\varphi_s$ into $\varphi _s = \displaystyle \sum_{k=0}^m s^k \varphi_{m-k}$. Since $\varphi_{m-k}$ corresponds to $(m-k ,m-k )$-form, we can compute $e^{- \lambda F_A} \varphi_s e^{\lambda F_A}{\cal O}_{\omega}$.
\begin{align}
&e^{- \lambda F_A} s^k \varphi_{m-k} e^{\lambda F_A}{\cal O}_{\omega} =e^{-2m \lambda } (s e^{2\lambda })^k \varphi _{m-k}{\cal O}_{\omega}.
\end{align}
Hence we obtain
\begin{align}
&e^{- \lambda F_A} \varphi_s e^{\lambda F_A} = e^{-2m\lambda }\varphi_{se^{2\lambda}}.
\label{conj}
\end{align}
Let us introduce vacuum vector $|0>$ and its dual $<0|$.  
Then we can represent the correlation function $<\varphi_{s} >$ as $<0|\varphi_{s} |0>$.
By using the relation (\ref{conj}), we obtain 
\ban
<\varphi_{s} > &=&<0|\varphi_s |0>=<0|e^{\lambda F_A} e^{-\lambda F_A} \varphi_s e^{\lambda F_A } e^{-\lambda F_A} |0>\\
&=&<0|e^{\lambda F_A} \varphi_{s e^{2\lambda}} e^{-\lambda F_A} |0>e^{-2m\lambda}.
\ean
Since our theory has $2m$ fermion zero modes $\chi^{i}_{0}$ and $\chi^{\bar{i}}_{0}$ $(i=1,\cdots,m)$, it is anomalous. Therefore, if we assign $|0>$ charge $(0,0)$, we have to assign $<0|$ charge $(m,m)$. Therefore, we have $F_{A}|0>=0$ and $<0|F_{A}=2m<0|$. Hence 
we obtain
\ban
<\varphi_{s} > &=&<0|e^{\lambda F_A} \varphi_{s e^{2\lambda}} e^{-\lambda F_A} |0>e^{-2m\lambda}\\
                   &=&e^{2m\lambda}<0| \varphi_{s e^{2\lambda}} |0>e^{-2m\lambda}\\
                   &=&<0| \varphi_{s e^{2\lambda}} |0>\\
                   &=&<\varphi_{s e^{2\lambda}}>.
\ean
This completes proof of the proposition.

\end{document}